\documentclass[11pt]{amsart}

\PassOptionsToPackage{pdfusetitle}{hyperref}
\PassOptionsToPackage{dvipsnames}{xcolor}

\usepackage[utf8]{inputenc}
\usepackage{lmodern}
\usepackage{mathtools,amsfonts,amsthm,mathrsfs,amssymb}

\usepackage[T1]{fontenc}

\usepackage{textcomp}
\mathtoolsset{showonlyrefs}

\usepackage[margin=1in]{geometry}

\usepackage[shortlabels]{enumitem}

\usepackage{tikz}
\usetikzlibrary{backgrounds}
\usepackage{algorithm,algpseudocode}
\usepackage{subcaption}
\usepackage[shortlabels]{enumitem}
\usepackage[dvipsnames]{xcolor}
\usepackage{multirow,hhline}
\usepackage{booktabs}
\usepackage{url}

\usepackage{graphicx}
\graphicspath{{./figures}}

\usepackage{bookmark}
\usepackage{hypcap}
\usepackage{microtype}

\theoremstyle{definition}

\newtheorem*{problem}{Problem}

\renewcommand{\epsilon}{\varepsilon}

\newlist{enumerata}{enumerate}{1}
\setlist[enumerata]{label=\upshape{(\alph*)}}
\setlist[enumerate]{label=\upshape{(\roman*)}}

\renewcommand{\Return}[0]{\State\textbf{return}\ }
\newcommand{\Select}[0]{\State\textbf{select}}
\renewcommand{\Comment}[1]{\hfill\textit{\# #1}}

\newcommand{\D}[0]{\mathcal{D}}
\newcommand{\districtof}[1]{\D(#1)}
\newcommand{\cutedges}[2]{\partial_{#1} #2}
\newcommand{\inducedsubgraph}[2]{#1[#2]}
\newcommand{\fiedler}[1]{\boldsymbol{f}_{#1}}

\title{Spectral partitioning of graphs into compact, connected regions}
\date{\today}

\author[E. Davies]{Ewan Davies}
\address{Department of Computer Science, Colorado State University, Fort Collins, USA}
\email{research@ewandavies.org}

\author[R. Job]{Ryan Job}
\address{Department of Computer Science, Colorado State University, Fort Collins, USA}
\email{Ryan.Job@colostate.edu}

\author[M. Kampbell]{Maxine Kampbell}
\address{Department of Computer Science, Colorado State University, Fort Collins, USA}
\email{maxinekampbell@gmail.com}

\author[H. Kim]{Hannah Kim}
\address{Department of Computer Science, Colorado State University, Fort Collins, USA}
\email{Gayeon.Kim@colostate.edu}

\author[H. Seo]{Hyojin Seo}
\address{Department of Computer Science, Colorado State University, Fort Collins, USA}
\email{Hyojin.Seo@colostate.edu}

\begin{document}
\begin{abstract}
We define and study a spectral recombination algorithm, SpecReCom, for partitioning a graph into a given number of connected parts. It is straightforward to introduce additional constraints such as the requirement that the weight (or number of vertices) in each part is approximately balanced, and we exemplify this by stating a variant, BalSpecReCom, of the SpecReCom algorithm. We provide empirical evidence that the algorithm achieves more compact partitions than alternatives such as RevReCom by studying a $56\times 56$ grid graph and a planar graph obtained from the state of Colorado.
\end{abstract}

\maketitle

\section{Introduction}

Clustering is one of the fundamental data processing tasks in statistics and machine learning. 
The main goal is to partition a dataset into so-called clusters such that data points belonging to the same cluster are more similar (in some domain-specific sense) to each other than to points from other clusters. 
We are interested in \emph{graph partitioning} which is a natural subclass of clustering problem where the dataset consists of a \emph{graph}: a set of points called vertices together with a set of (undirected) edges which connect pairs of points. 
The goal is to partition the set of vertices in some way that takes into account the spatial information contained in the edges. 
For example, one might like to partition the graph into clusters such that each cluster is more densely connected to itself than to any other cluster. 

In physics and data science, one natural way that undirected graphs arise is as a notion of locality. We consider a set of points for which we have some concept of proximity or adjacency, and put an edge between each ``close'' pair of points. 
Lattice graphs are used in physics  to represent statistical models of systems with inter-particle forces where the interactions are simplified to act only along edges of the graph. 
A planar map such as the electoral precincts of a state in the USA can be turned into a graph where each precinct is a vertex and an edge connects each pair of precincts that share a border. 
We present an algorithm suited to partitioning graphs of this type, which are said to be the \emph{dual graph} of a planar map.

Our main contribution is a flexible algorithm ``SpecReCom'' for graph partitioning subject to the constraint that each cluster forms a connected subgraph, with the provision for additional constraints such as the requirement that the clusters have equal size. 
This is common requirement in applications, and we present ``BalSpecRecom'' to handle this case specifically.
We combine two themes in graph partitioning that each have several important applications: spectral partitioning and recombination. 
Spectral partitioning is a graph partitioning technique which uses spectral properties of the adjacency matrix (or more commonly the Laplacian matrix) of a graph to cluster the vertices (see e.g.~\cite{fiedler_connectivity,chung_sgt}), and recombination is a stochastic technique for producing a sequence of graph partitions by combining and splitting parts of the partition~\cite{deford_recomfamily}. 
Our main application is partitioning of dual graphs of planar maps, and our algorithm is designed to produce a partition such that each part induces a connected subgraph and such that each part is highly ``compact'' in a way that we make precise below. 
We discuss an application of this algorithm to a problem in political science in Section~\ref{sec:politics}.

\subsection{Connected graph partitions}\label{sec:partitions}

A \emph{graph} $G=(V,E)$ is a set $V$ of vertices (called nodes in some works) together with a set $E$ of unordered pairs from $V$. 
A \emph{partition} of the graph $G=(V,E)$ is a set $\mathcal D = \{D_1,\dotsc,D_k\}$ such that each part $D_i$ is a subset of $V$ (written $D_i\subset V$), the union of the parts is $V$ (written $\bigcup_{i=1}^k D_i=V$), and such that the parts are pairwise disjoint (for all $i\ne j$ we have $D_i\cap D_j=\emptyset$).
We are specifically interested in \emph{connected} partitions of $G$, which are partitions of $G$ with the additional requirement that each part $D_i$ induces a connected subgraph of $G$.
When we want to be specific about the number $k$ of parts, we call the desired partition a \emph{connected $k$-partition}.
We consider the following graph partitioning problem. 

\begin{problem}
Given a graph $G=(V,E)$, a positive integer $k$, and a set $C$ of additional constraints, create a connected $k$-partition $\mathcal D = \{D_1,\dotsc, D_k\}$ of $G$ such that the additional constraints in $C$ are satisfied.  
\end{problem}

Our algorithm can be applied in the case that there are no additional constraints, though we also demonstrate that it can be adapted to the case that we have a function $w:V\to [0,\infty)$ assigning nonnegative weights the vertices of $G$, and an additional constraint stipulating that the weights $w(D_i) = \sum_{v\in D_i} w(v)$ are close together.

Without further specification, in the case where there are no additional constraints it is not too hard to devise efficient algorithms that solve the above problem. 
In many applications, however, one is interested in a very small subset of the connected $k$-partitions that maximize some notion of ``compactness'' of the parts. 
This is akin to the situation in the broader clustering literature: there are many simple ways to divide datasets into $k$ clusters, but in most applications we want a clustering that optimizes some objective, for example a $k$-means clustering optimizes an objective function known as the ``within-cluster sum of squares'' (which we do not dwell on here). 

Our algorithm is designed to heuristically optimize a notion of compactness called ``cut-edge count'', defined as follows. 
Given a partition $\mathcal D$ of a graph $G=(V,E)$, we say that an edge $uv\in E$ is \emph{cut} by $\mathcal D$ if $u$ and $v$ lie in different parts of the partition.
We are interested in \emph{minimizing} the number of cut edges of our $k$-partitions as there is evidence that when partitioning the dual graph of a planar map, the number of cut edges correlates with other more geometric notions of compactness coming from the geometry of the planar map itself~\cite{clelland_recom}. 
If we dispense with the requirement that our $k$-partitions are connected then this is actually a well-studied problem in computer science called ``minimum $k$-cut''. 
The case $k=2$ is rather famous and has a polynomial-time algorithm due to Karger (see~\cite{Kar93,KS96a}), though subject to the constraint that the parts of the partition have equal size the problem is NP-hard and known as ``minimum bisection'' (see e.g.~\cite{GJ09}).
Minimizing the number of cut edges subject to a connectivity constraint underpins a relatively new direction in graph partitioning (see e.g.\ the survey~\cite{deford_recomfamily}, although there optimization is not the primary goal).

Our SpecReCom algorithm fits into a family of partitioning algorithms known as \emph{recombination}, where a random process (a Markov chain) is used to generate a sequence of graph partitions by starting with a given partition and modifying it step-by-step with steps that combine two parts into one super-part and then split the super-part back into two parts. 
Though this sounds like a simple idea, it is exceptionally useful when one has a number of intricate constraints to satisfy such as connectedness and that parts have equal size. 
The recombination setup allows one to focus on splitting graphs into just two parts subject to these constraints rather than directly solving the $k$-partitioning problem. 
The early work on recombination focused on a specific political application that we describe in the next subsection.

The original recombination method~\cite{deford_recomfamily} takes a connected $k$-partition of a graph and performs a ``step'' which randomly combines two adjacent parts into one super-part (which one can show is also connected). Then, one splits the super-part into two smaller parts by generating a random spanning tree of the super-part and finding whether there is an edge one can cut in that tree to produce two connected parts such that ones additional constraints are satisfied.

We broaden the existing understanding of recombination methods by proposing an alternative way to split the super-part into two connected pieces based on spectral graph theory. 
Investigating this idea was proposed by DeFord~\cite{deford_personal}, one of the original authors of the spanning tree recombination method.
The key idea of SpecReCom is to split the super-part into two based on the standard notion of spectral partitioning developed in 1973 by Fiedler~\cite{fiedler_connectivity}. 
This notion is often encountered in graph theory by way of \emph{Cheeger's inequality} (see e.g.~\cite{AM85,Alo86,SJ89}) which relates eigenvalues of the adjacency matrix (or Laplacian matrix) of a graph to properties of partitions of the graph into two parts. 

There is a rich literature on spectral partitioning in graph theory, though it is unusual to focus on connectivity in this domain. 
Notably, Urschel and Zikatanov~\cite{urschel_bisection} refined the work of Fiedler and studied the robustness of Fiedler's method to partition a graph into two \emph{connected} subgraphs. 
Though it is challenging to extend spectral partitioning ideas into the case of more than two parts, significant work has gone in that direction~\cite{OT12,lee_cheegercut,KLL+13,OT14}. 
The sophistication of these techniques and the guarantees that one can prove are very impressive, but it seems extremely challenging to address the constraints of connectivity and (approximately) equal part-size (or more generally equal weights subject to a weight function on the vertices) in those frameworks. 
Thus, we found it natural to combine the simpler case of spectral partitioning into two parts with the recombination technique that lets us solve the partitioning problem for larger numbers of parts. 

We extend the standard spectral partitioning idea with what one might call a ``Cheeger sweep'' to efficiently handle the additional constraint of balanced vertex weights in the parts, though there are some complexities that arise when one wants both connectedness and balanced vertex weights. That is, we define another algorithm BalSpecReCom which is an extension of SpecReCom that attempts to balance the vertex weights in the parts.
We present some statistical analysis that demonstrates how BalSpecReCom does achieve its aims on representative data.

\subsection{Political districting}\label{sec:politics}

Finding connected $k$-partitions of the dual graph of a planar map is highly relevant to representative democracies, where some territory is divided into regions for the purpose of electing political leaders. 
A common case, seen at several different scales in the USA and UK for example, is that a territory composed of small, indivisible ``units'' (variously known as precincts or wards) is divided into a few regions by grouping neighboring units into districts (or constituencies). 
In many political systems, each district elects a single political representative by pooling the votes from the people registered in the units comprising the district in a winner-takes-all election (known as first-past-the-post or plurality voting). 
In this case, subject to fixed votes for political parties in each unit, the location of the district boundaries can have a striking effect on the outcomes of the election (see e.g.\ the demonstrations in~\cite{duchin_politicalgeometry}). 
The boundaries themselves are subject to a number of constraints, most often including connectivity, and various political processes exist to allow for the drawing and redrawing of boundaries subject to such constraints. 
Note that a basic requirement for fairness is that each of the parts of the $k$-partition contains (approximately) equal (voting) population, as otherwise a vote in one district could be more influential than a vote in another. 
We call the process of partitioning planar maps in this way \emph{political districting}, and the connected $k$-partitions produced to meet this end are called \emph{districting plans}. 

Several authors (e.g.~\cite{For64,vickery_gerrymandering} to cite early examples from the 1960s) have called for the study of algorithms that generate districting plans. 
One of the major concerns is that incumbent political entities can use their powers to redraw the district boundaries in ways that favor their interests in the elections. 
This has a long history, including developments in the USA that led to the coining of the term \emph{gerrymandering} in early 1800s to describe such manipulation of the boundaries. 
One of the key problems one faces in assessing whether gerrymandering is occurring or to what degree some proposed changes to district boundaries is fair, is that there is no obvious choice for a ``non-gerrymandered'' baseline to compare to. 
Another problem is the precise definition of fairness one should work with, which a rich and complex problem about which we say nothing in this work. 

An idea that appears again and again in the history of political districting is that ``manipulated'' boundaries are winding and snakelike (or salamander-like), and that ``natural'' boundaries yield so-called ``compact'' districts. Defining any of these terms precisely is not a straightforward task, and there are a number of geometric and graph-theoretic measures of compactness that are in use. We refer the reader to discussions in the literature~\cite{duchin_politicalgeometry,clelland_recom} of such scores, and confine our interest to the relatively new measure of cut edge count described in the previous section that has been considered successful in recent studies. 
As well as the intrinsic interest in minimum $k$-cut from the computer science community, the desire for compact districting plans informs the design of our main contribution SpecReCom.

In the last decade, there has been a steady growth of interest in computational techniques that attempt to solve the problem of detecting gerrymandering by producing ``ensembles'' of very large numbers of districting plans in a purportedly unbiased manner~\cite{chen_detectgerrymander,cohenaddad_polygondistricts,deford_recomfamily}. Then one can build a statistical picture of what the outcome of elections would be with a non-gerrymandered plan. This is known as \emph{ensemble-based analysis}. Of course one could attempt to subvert this process by designing an algorithm that favors the production of plans that suit ones political aims, but we posit that it is easier to study a well-specified algorithm for bias than it is to scrutinize opaque (human) decision-making. 
As such one can consider our contribution of novel techniques for graph partitioning subject to constraints such as connectivity and population balance as useful in political districting.

\subsection{Outline}

In Section~\ref{sec:related} we discuss related works, in Section~\ref{sec:background} we present the technical background that underpins SpecReCom and BalSpecReCom, in section~\ref{sec:specrecom} we specify these two algorithms in detail, in section~\ref{sec:implementation} we briefly describe our implementation of them, in section~\ref{sec:evaluation} we present and discuss our experimental results, in section~\ref{sec:conclusions} we provide several conclusions based on our experiments, and finally in section~\ref{sec:future} we discuss potential avenues of future research.

\section{Related Work} \label{sec:related}

\subsection{Spectral partitioning}

    Spectral methods for partitioning graphs into more than two parts were studied with various coauthors by Oveis Gharan and Trevisan~\cite{OT12,lee_cheegercut,KLL+13,OT14}. 
    Our work can be seen as an attempt to unite this thread of research with the recombination literature, and more broadly to study graph partitioning subject to constraints that are required in the political districting problem.
    
    A result of Lee, Oveis Gharan and Trevisan~\cite{lee_cheegercut} states that a graph has a partition into $k$ parts with few cut edges if and only if there are $k$ eigenvalues in the spectrum of its Laplacian matrix that are close to zero.
    Their general-purpose algorithm works by embedding each vertex of the graph in a real vector space, then outputting subsets of vertices that form least-expanding sets.
    While this technique does produce a $k$-partition with small cut edge count, and can be modified to take into account the need to balance vertex weights in the parts, it does not appear suited to connectivity in the context of the dual graphs of planar maps. 
    For such graphs we may not have control over the graph spectrum and the choice of the number $k$ of parts is usually extrinsic (related to, e.g.\ population in the case of districting), not driven by its spectral properties.

    We note that in the case of graphs with vertex weights, which is highly relevant to some applications of graph partitioning, there is a notion of graph Laplacian due to Chung and Langlands~\cite{chung_vertexlaplacian} which one might want to use. 
    Extensive testing showed that spectral partitioning with this Laplacian performed significantly worse than with the standard Laplacian from the perspective of connectedness. 
    Since our main application has connectedness as a hard requirement and weight-balance as a soft requirement, we were forced to optimize for weight balance a different way. 
    This led to the adoption of a so-called ``Cheeger sweep'' in SpecReCom, see the algorithm we call BalSpecReCom.
    
\subsection{Political districting}

    Chen and Rodden~\cite{chen_detectgerrymander} published one of the early works on the idea of districting plan comparison in 2015. With the intent to detect gerrymandered districting plans, they focused on generating a benchmark plan to compare currently implemented plans against. In essence, their algorithm works by starting with each unit as its own district, and whittling down the number of districts by combining two at a time. Then, the algorithm repeatedly flips units district membership along boundaries to improve population balance.
    
    In 2017, Kawahara et al.\ \cite{kawahara_disparity} described an algorithm for exhaustively enumerating graph partitions for districting plans. They focus on using a so-caled ``zero-suppressed decision diagram'' data structure to explore the space of all districting plans given a graph and district count. They also detail how to modify the original algorithm to consider population disparity, making sure that only plans with bounded disparity are enumerated.
    
    Cohen-Addad et al.\ \cite{cohenaddad_polygondistricts} provided another algorithm for creating districting plans utilizing power diagrams in 2018. Their algorithm uses centroidal power diagrams to determine the districts to which population units should be assigned, by solving the balanced k-means clustering problem. These districts end up being polygons whose centroids define the center points of each power region in the diagram. Notably, the average number of sides for all polygons is less than 6 and their populations differ by at most 1, and so the districts end up being balanced and visually compact.
    
    Levin and Friedler \cite{levin_voronoi} in 2019 used a similar idea with Voronoi diagrams. Like Cohen-Addad et al, their algorithm iterates using two components, a Voronoi component to bipartition regions and a swapping component to ensure population balance and connected districts. But in contrast, Levin and Friedler use a divide-and-conquer algorithm, bipartitioning the entire state first then recursively dividing unitl the desired number of parts is reached. 

    Fifield et al.\ \cite{fifield_firstmcmc} proposed the first redistricting algorithm based on Markov chain Monte Carlo (MCMC) methods in 2020.
    MCMC methods are widely applied in the sciences to study complex and high-dimensional datasets, see the next section for details. They are particularly suited to producing districting plans as they naturally allow one to sample a large number of plans from a probability distribution over plans, and indeed a large part of working with MCMC methods is carefully designing an algorithm that is both efficient and whose outputs are from a desirable distribution.
    The basic idea for a MCMC method for generating districting plans is to start with a districting plan and define a randomized process for how to take a ``step'' to another districting plan.
    A step of the Markov chain designed by Fifield et al.\ randomly selects a few units along the current district boundaries and then flips the district assignment of those units. This new plan can then be accepted or rejected (probabilistically) before continuing to the next step.
    
    Ensemble-based analysis, the idea to generate an ensemble of districting plans for comparison with a proposed real-world plan, was enacted by Herschlag et al.\ \cite{herschlag_applyensembles}, also in 2020. Building on Fifield et al.'s work, they use their MCMC method with an added simulated annealing procedure to generate this ensemble of plans.
    
    In 2021, Clelland et al.\ published on the ReCom MCMC method developed by the Data and Democracy Lab in 2018, and DeFord et al.\ characterized it as one method among a family of recombination methods \cite{clelland_recom, mggg_techreport, deford_recomfamily}. 
    The basic idea of ReCom is that one can step to a new districting plan by combining two adjacent districts into one super-district and then randomly split the super-district back into connected pieces (e.g.\ subject to population balance). 
    The original ReCom chain splits districts by generating random a spanning tree of the super-district and cuting an edge of it to produce two connected districts. 
    These groups introduce the idea that considering probability distributions on districting plans which weight them according to how many spanning trees each part admits is important for compactness in districting plans, and they show how to target this distribution using MCMC methods.
    
    In 2022, Cannon et al.\ \cite{cannon_revrecom} modified the ReCom method to satisfy a property known as \emph{reversibility}, resulting in RevReCom. Reversibility is an important step towards being able to prove guarantees about the output of the algorithm, and through showing reversibility Cannon et al.\ ensure that RevReCom converges to a precise ``spanning tree distribution''. They also developed a high-quality implementation of their algorithm for real-world use~\cite{rust_revrecom}.
    

\section{Technical background}\label{sec:background}

    


\subsection{Markov chain Monte Carlo}

    For the purposes of applying ensemble-based analysis, one can efficiently sample a districting plan from a space of valid plans using Markov chains. The choice of chain influences which types of plans are sampled. This technique is known as Markov chain Monte Carlo (MCMC).
    
    Starting from an initial plan $\sigma_0$, a new random plan is sampled by starting from an initial plan $\sigma_0$ and running a process which makes discrete ``steps'' by proposing (at random) a modification of the current plan and then deciding whether to accept the proposed change. 
    This process is called a Markov chain when the \emph{proposal function} $p$ satisfies certain properties\footnote{We do not go into the details here, see e.g.~\cite{duchin_politicalgeometry} for a gentle introduction in the setting of our main application.}, and is usually run for a predetermined number of steps $N$.

    The library ``GerryChain'' \cite{gerrychain}, created in 2018 by the Data and Democracy Lab (formerly the MGGG Redistricting Lab), provides a framework for implementing such MCMC methods in Python. A simplified version of GerryChain's generic MCMC method is shown in algorithm~\ref{alg:mcmcredist}.

    \begin{algorithm}[htbp]
        \caption{Markov chain Monte Carlo Redistricting}
        \begin{algorithmic}[1]
            \Procedure{Redistricting}{$N, C, \sigma_0, p$} \Comment{input: steps, constraints, initial state, proposal}
                \For{$1 \leq i \leq N$}
                    \Repeat
                        \State $\sigma_i \gets p(\sigma_{i-1})$
                    \Until{$\sigma_i$ is feasible under $C$} \Comment{don't step out of the state space}
                \EndFor
                \Return $\sigma_N$ \Comment{output: final state}
            \EndProcedure
        \end{algorithmic}
        \label{alg:mcmcredist}
    \end{algorithm}

\begin{figure}[htbp]
        \centering
        \begin{subfigure}[c]{0.25\linewidth}
            \includegraphics[width=\linewidth]{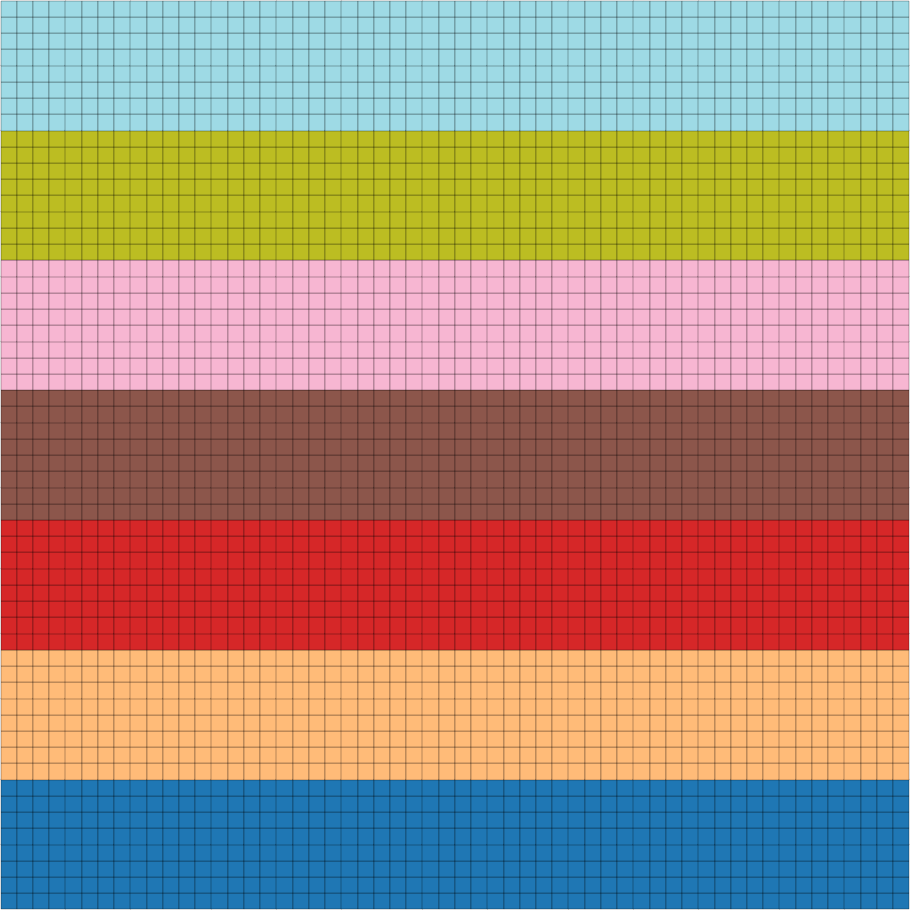}
            \caption{A $56\times 56$ grid graph partitioned into seven connected parts.}
            \label{fig:initialgrid}
        \end{subfigure}
        \hfill
        $\displaystyle \xrightarrow[\text{\makebox[\widthof{ecombin}][c]{recombine}}]{}$
        \hfill
        \begin{subfigure}[c]{0.25\linewidth}
            \includegraphics[width=\linewidth]{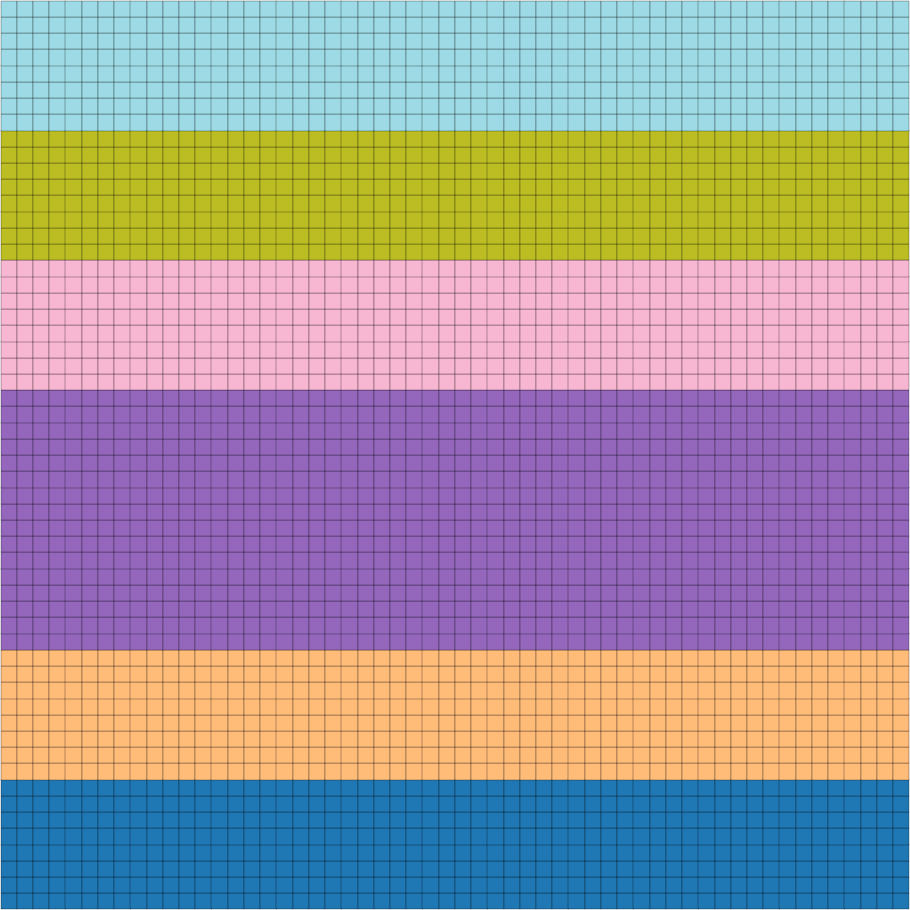}
            \caption{The same graph after recombining the brown and red districts.}
        \end{subfigure}
        \hfill
        $\displaystyle \xrightarrow[\text{\makebox[\widthof{ecombin}][c]{split}}]{}$
        \hfill
        \begin{subfigure}[c]{0.25\linewidth}
            \includegraphics[width=\linewidth]{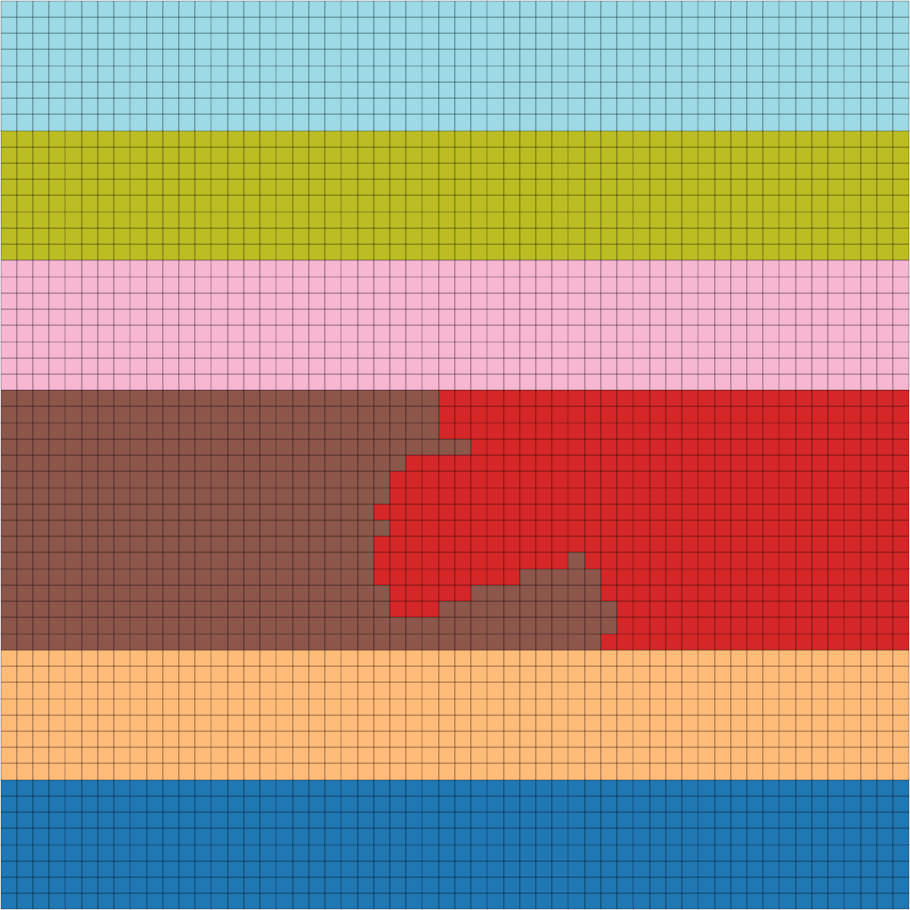}
            \caption{The graph after splitting the super-district back into two districts.}
        \end{subfigure}
        \caption{An example step of a recombination algorithm on a $56 \times 56$ grid graph.}
        \label{fig:recom}
    \end{figure}

    Our contributions are part of a family of MCMC methods studied in the literature with \emph{recombination} proposals. These types of algorithms follow the same basic structure (see Figure~\ref{fig:recom}):
    \begin{enumerate}[1.]
        \item Select two adjacent parts in the current $k$-partition
        \item Recombine the two parts into one (connected) super-part
        \item Split the combined super-part back into two parts
    \end{enumerate}

\subsection{Recombination}
    The first recombination chain (ReCom) \cite{mggg_techreport,deford_recomfamily} splits the combined parts by sampling a uniform random spanning and determining whether there is an edge of it which can be cut to yield a valid $k$-partition.

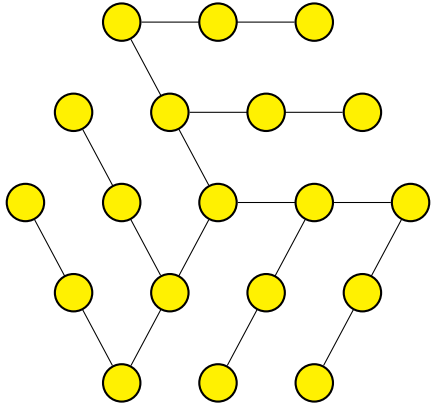
\begin{figure}[htbp]
        \centering
        \begin{tikzpicture}[scale=0.8,thick, state/.style={draw, circle, inner sep=5pt, minimum size=12pt, fill=yellow}] 
            \foreach \n [count=\i from -1] in {1,2,3} {
                \node [state] (1\n) at (\i*16mm, 0) {};
                \node [state] (5\n) at (\i*16mm, 4*15mm) {};
            }
            \foreach \n [count=\i from -2] in {1,2,3,4} {
                \node [state] (2\n) at (\i*16mm+8mm, 15mm) {};
                \node [state] (4\n) at (\i*16mm+8mm, 3*15mm) {};
            }
            \foreach \n [count=\i from -2] in {1,2,3,4,5} {
                \node [state] (3\n) at (\i*16mm, 2*15mm) {};
            }

            \begin{scope}[on background layer]
                \draw (11) -- (33);
                \draw (12) -- (34);
                \draw (13) -- (35);
                
                \draw (51) -- (53);
                \draw (42) -- (44);
                \draw (33) -- (35);
                
                \draw (51) -- (33);
                \draw (41) -- (22);
                \draw (31) -- (11);
            \end{scope}
        \end{tikzpicture}
        \caption{A spanning tree with no $\varepsilon$-balance edges for $\varepsilon < 7/19$.}
        \label{fig:nobaledgetree}
    \end{figure}

    This method works in a setting with vertex weights and  a constraint that the parts are (approximately) balanced in weight. In order to maintain balanced part weights, the algorithm cannot select an edge of the spanning tree arbitrarily.
    Instead, after an edge is sampled uniformly, it ensures that the difference in weight between the two parts is limited by a tolerance parameter. But for any given spanning tree, it is not guaranteed that there exists an edge that satisfies the population tolerance when cut (see Figure~\ref{fig:nobaledgetree}). So in this case, the algorithm rejects the tree and samples a new one.
    
    Because ReCom samples spanning trees in determining how to split the recombined parts, one might expect that after running the chain for a large number of steps, the distribution of the $k$-partition approximates the ``spanning tree'' distribution, as described in~\cite{clelland_recom,cannon_revrecom}. 
    This selects a spanning $k$-partition of a graph with probability proportional to the product of the number of spanning trees each part admits.
    A refinement of the original ReCom algorithm known as Reversible Recombination (RevReCom)~\cite{cannon_revrecom} was designed so that its stationary distribution is precisely the spanning tree distribution. 

    Similar to ReCom, RevReCom functions by sampling a random spanning tree of the recombined super-part's graph, and then cutting an edge that maintains some weight tolerance. This tolerance is characterized by ``$\varepsilon$-balance edges'', edges of the spanning tree that, when cut, maintain a weight tolerance across the whole graph $G$ of at most $\varepsilon$.

    RevReCom, however, increases the probability that the proposed split is rejected. A parameter $m$ needs to be chosen that is at least the absolute maximum number of $\varepsilon$-balance edges that exist for any spanning tree of any two parts of the graph. This value is impractical to calculate, and so it is often determined empirically~\cite{cannon_revrecom}. Choosing a value of $m$ that is too large causes many more rejections than necessary, since the likelihood that RevReCom rejects the proposed cut is $1 - b/m$ where $b$ is the number of $\varepsilon$-balance edges.

    Moreover, RevReCom can also reject the split based on how many edges it would cut in $G$. The overall increased likelihood of rejection means that RevReCom runs more slowly than ReCom, as it spends more time rejecting splits. This problem was mitigated in implementation by allowing many different spanning trees to be considered in parallel, and moving on once one of them is accepted~\cite{rust_revrecom}.

\subsection{Spectral partitioning and Fiedler vectors}
    Based on methods from the literature \cite{deford_recomfamily,clelland_recom,cannon_revrecom}, we present a new type of recombination algorithm that uses a Fiedler vector \cite{fiedler_connectivity} of the recombined super-part to determine how to split it up.
    
    For an undirected, finite, simple graph $G$ with $n$ vertices, we consider the \emph{Laplacian} of $G$, $L = D - A$, where $D$ and $A$ are the degree and adjacency matrices of $G$, respectively. 
    That is, $D$ is a diagonal $n\times n$ matrix indexed by the vertices of $G$ such that $D_{v,v}$ is the degree of $v$ in $G$ and $A$ is a symmetric $n\times n$ matrix indexed by the vertices of $G$ such that $A_{u,v} = 1$ if $uv\in E(G)$ and $A_{u,v}=0$ otherwise.
    For graphs with weighted edges, one can put the weight of the edge $uv$ into the entry $A_{u,v}$, and we turn to this idea in the implementation of our algorithm.
    
    For our work, we leave the Laplacian un-normalized, but in some contexts~\cite{chung_sgt}, the \emph{normalized Laplacian} matrix $\mathcal{L} = D^{-1/2} L D^{-1/2}$ is used.

    The spectrum of the graph $G$ is the set of eigenvalues of $L$, all of which are nonnegative and exactly one of which is zero for connected graphs~\cite{chung_sgt}. Listing the spectrum in order, we have:
    \[
        0 = \lambda_1 \leq \lambda_2 \leq \cdots \leq \lambda_n,
    \]
    where $\lambda_2$ is referred to as the algebraic connectivity of $G$. 
    An eigenvector with eigenvalue $\lambda_2$ is called a \emph{Fiedler vector} of $G$, $\fiedler{G}$. While it is possible that the geometric multiplicity of $\lambda_2$ is greater than 1, corresponding to meaningfully different choices of Fiedler vector, we do not explore this possibility in this work. We discuss this further in section~\ref{sec:alternatefiedler}.

    Urshel and Zikatanov \cite{urschel_bisection} show that for any connected, undirected, simple graph $G$, there exists a Fiedler vector $\fiedler{G}$ such that the vertex bipartition given by 
    \[
        \{\fiedler{G}^+, \fiedler{G}^-\} = \big\{\{v \in V_G : \fiedler{G}[v] \geq 0\},\ \{v \in V_G : \fiedler{G}[v] < 0 \} \big\}
    \]
    can be used to induce two connected subgraphs of $G$: $\inducedsubgraph{G}{\fiedler{G}^+}$ and $\inducedsubgraph{G}{\fiedler{G}^-}$.

    Urshel and Zikatanov's results on connectivity provide motivation for considering the approach in congressional redistricting. The bipartition described above makes use of the negative and non-negative entries in the Fiedler vector. Notably, cut-vertices of graphs have corresponding entries of zero, and so other vertices can be thought of as being on either side of this cut. The intuition behind spectral bipartitioning is demonstrated in figure~\ref{fig:spectral_bipartition}, in which each vertex is labeled and colored to correspond with its Fiedler vector entry, rounded to the nearest thousandth.

\subsection{Further notation conventions}

    We list some notation used in the remainder of this paper. Let $\D = \{D_1, \cdots, D_k\}$ be a connected $k$-partition plan for a graph $G=(V,E)$, and let $v \in V$, $e \in E$, and $S \subseteq V$, then:
    \begin{itemize}
        \item $\districtof{v}$ is the unique part in $\D$ that contains $v$.
        \item $\cutedges{G}{\D}$ is the set of edges of $G$ that are cut by $\D$.
        \item $w(v)$ is the weight of the vertex $v$ (e.g.\ its population in the districting setting), $w(S)=\sum_{v\in S}w(v)$ is the weight of a subset $S$ of vertices of $G$.
        \item $w(e)$ is the weight of the edge $e$.
        \item $\inducedsubgraph{G}{S}$ is defined as the subgraph of $G$ induced by the vertex set $S$:
        \[
            \inducedsubgraph{G}{S} = (S, \{uv \in E : u \in S \text{ and } v \in S\})
        \]
        \item We write $V(G) = V$ for the vertex set of $G$ and $E(G) = E$ for the edge set of $G$.
        \item In the districting setting, the \emph{population deviation} of $\D$ is the furthest any individual part $D_i \in \D$ is from the ideal population, $w(V)/k$. It is defined as:
        \[
            \mathrm{popDev}(\D, V) := \max_{D_i \in \D} \left| \frac{k \cdot w(D_i)}{w(V)} - 1 \right|.
        \]
    \end{itemize}

\section{The Spectral Recombination Proposal} \label{sec:specrecom}

We provide two algorithms that build on the ideas present in ReCom \cite{clelland_recom} and RevReCom \cite{cannon_revrecom} from the literature. Our first algorithm, Spectral Recombination (SpecReCom), utilizes spectral bipartitioning as a step in recombination. The second algorithm, Balanced Spectral Recombination (BalSpecRecom), illustrates how Spectral Recombination can be modified to account for balancing the weight among parts.

\subsection{Spectral Recombination}
    Like ReCom, the Spectral Recombination (SpecReCom) proposal chooses two parts to recombine based on the number of edges between them, with part pairs that cut more edges of $G$ being more likely to be chosen. After the recombination step, the induced subgraph's Laplacian and Fiedler vector are computed, where each of the subgraph's vertices is associated with one of the entries in the Fiedler vector. SpecReCom splits the resulting super-part by assigning vertices with a negative entry to one part and those with a non-negative entry to the other. Algorithm~\ref{alg:specrecom} provides a precise description of SpecReCom.

    \begin{algorithm}[htbp]
        \caption{Spectral Recombination Proposal}
        \begin{algorithmic}[1]
            \Procedure{SpecReCom}{$\D; G$} \Comment{input: current state; graph}
                \Select\ $uv \in \cutedges{G}{\D}$ uniformly at random \Comment{sample a cut edge}
                \State $H \gets \inducedsubgraph{G}{\districtof{u} \cup \districtof{v}}$ \Comment{recombine}
                \For{$e \in E(H)$}
                    \State $w(e) \gets $ \Call{rand}{1,2} \Comment{randomize edge weights}
                \EndFor
                \State $H_1 \gets \inducedsubgraph{H}{\fiedler{H}^+}$ \Comment{apply spectral bipartitioning}
                \State $H_2 \gets \inducedsubgraph{H}{\fiedler{H}^-}$
                \State $\D' \gets (\D \setminus \{\districtof{u}, \districtof{v}\}) \cup \{V(H_1), V(H_2)\}$
                \State \textbf{return} $\D'$ \Comment{output: next state}
            \EndProcedure
        \end{algorithmic}
        \label{alg:specrecom}
    \end{algorithm}
    
\begin{figure}[htbp]
        \centering
        \begin{subfigure}[c]{\linewidth}
            \centering
            \includegraphics[width=0.45\linewidth]{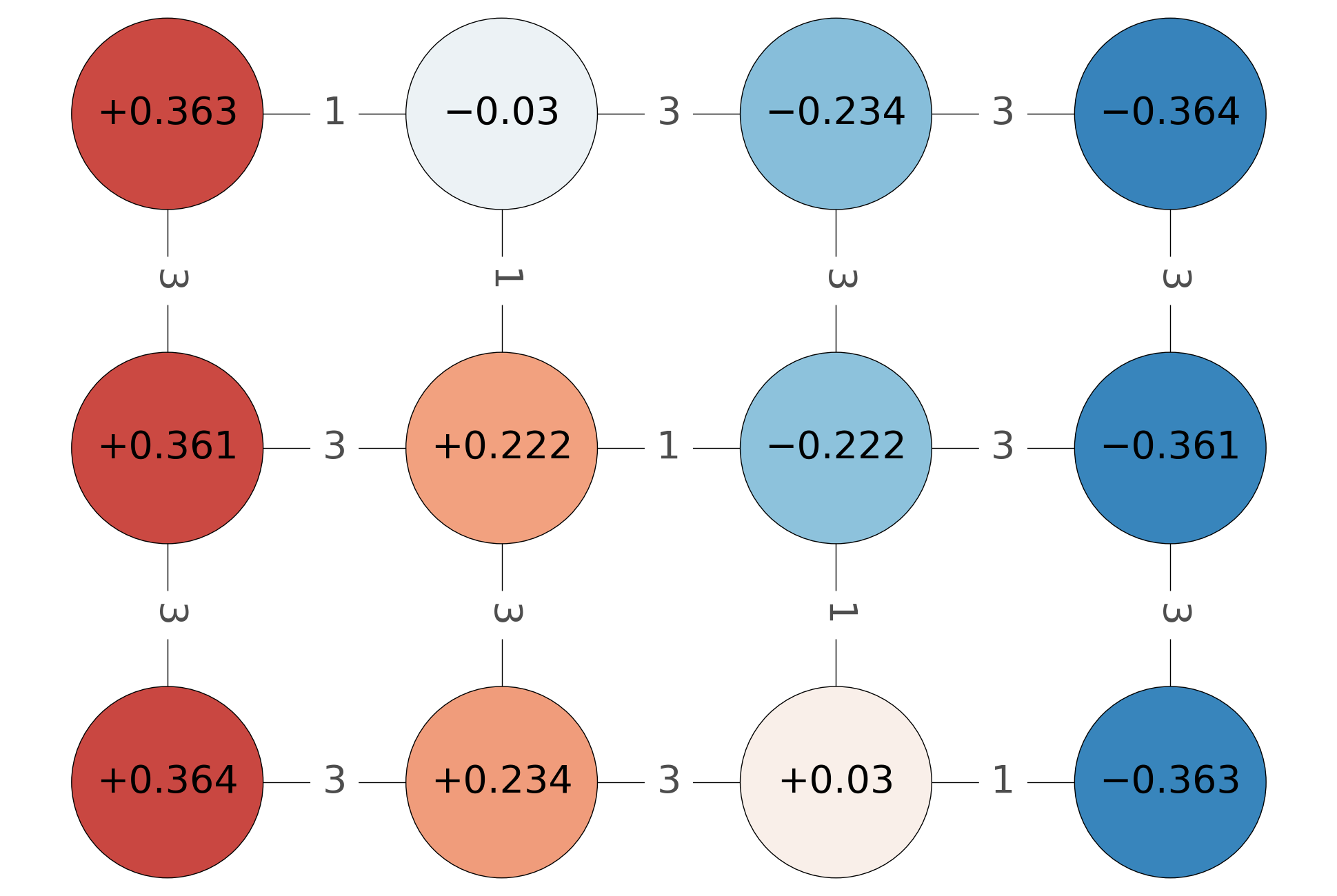}
            \caption{Edge weights determine how ``easy'' they are to cut.}
        \end{subfigure}
        \\\vspace{4ex}
        \begin{subfigure}[c]{0.45\linewidth}
            \centering
            \includegraphics[width=\linewidth]{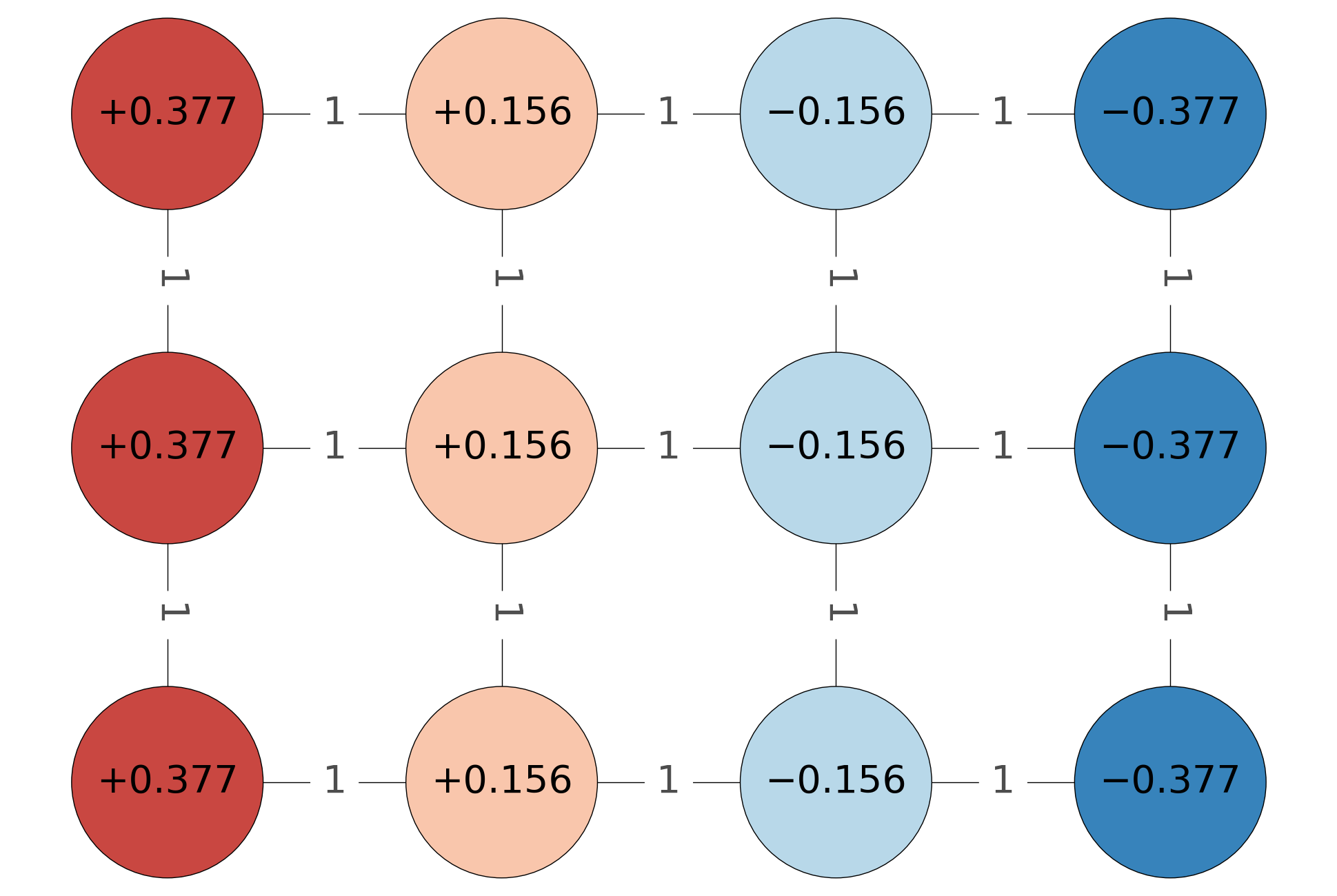}
            \caption{A graph with all edges having equal weight.}
        \end{subfigure}\hspace{0.08\linewidth}
        \begin{subfigure}[c]{0.45\linewidth}
            \centering
            \includegraphics[width=\linewidth]{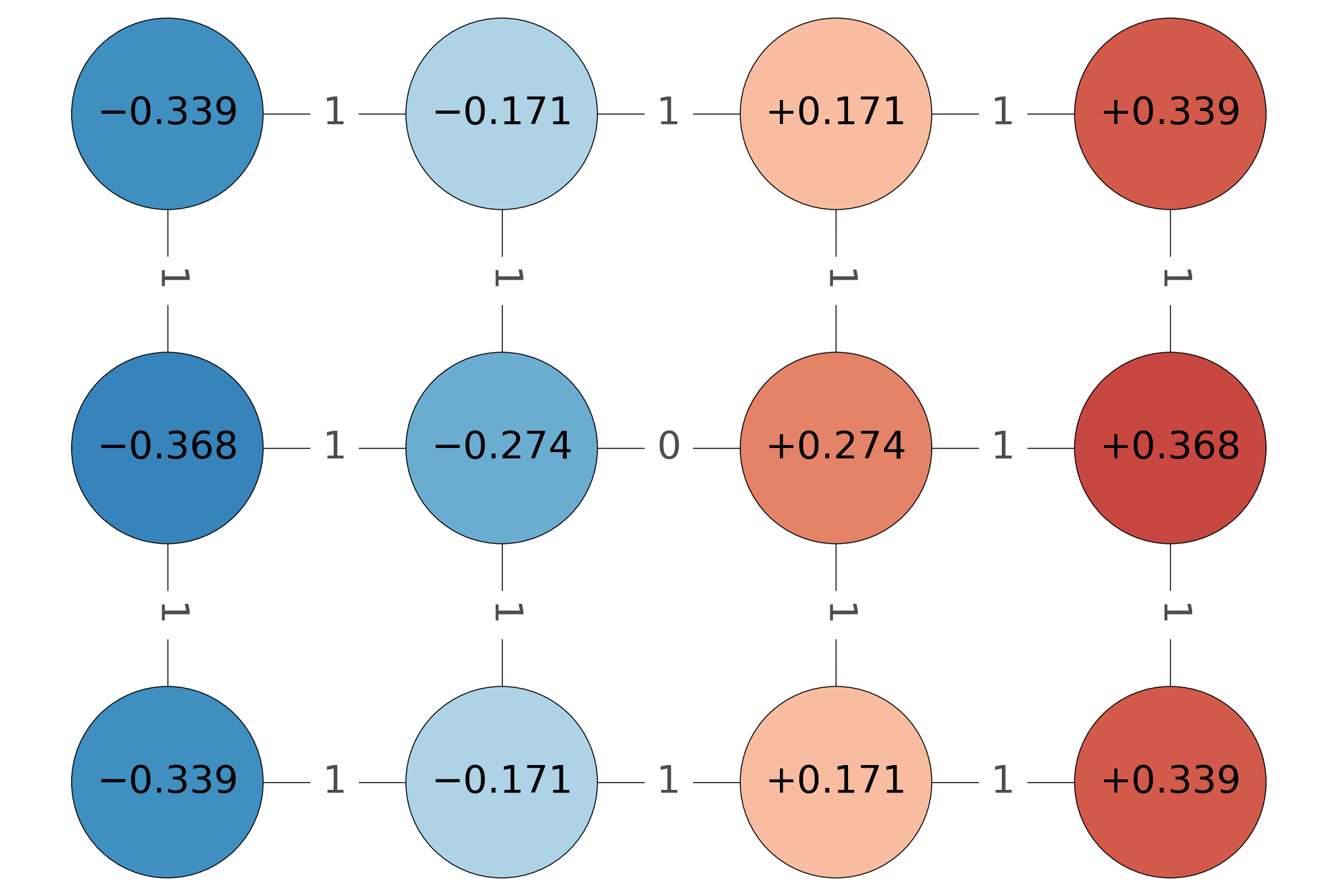}
            \caption{Edges with weight zero can be cut ``for free''.}
        \end{subfigure}
        \caption{How edge weights impact Fiedler vector entries.}
        \label{fig:spectral_bipartition}
    \end{figure}

    Lines 4-6 of Algorithm~\ref{alg:specrecom} show that, to introduce non-determinism into SpecReCom, we first randomize the edge weights of $G$ by making uniform choices in the interval $[1,2]$. For the purposes of bipartitioning via the Fiedler vector, edge weights influence which edges are more or less ``costly'' to cut. Without random edge weights, all edges would have equal cost and the proposal would be deterministic. Various edge weight configurations, along with the Fiedler vector entries they induce, are shown in figure~\ref{fig:spectral_bipartition}. Our choice to use weights between 1 and 2 was arbitrary, but it appears to provide sufficient variety in the generated connected $k$-partitions that the algorithm did not get ``stuck'' in our empirical analysis.

\begin{figure}[htbp]
        \centering
        \begin{subfigure}[c]{0.42\linewidth}
            \centering
            \includegraphics[width=\linewidth]{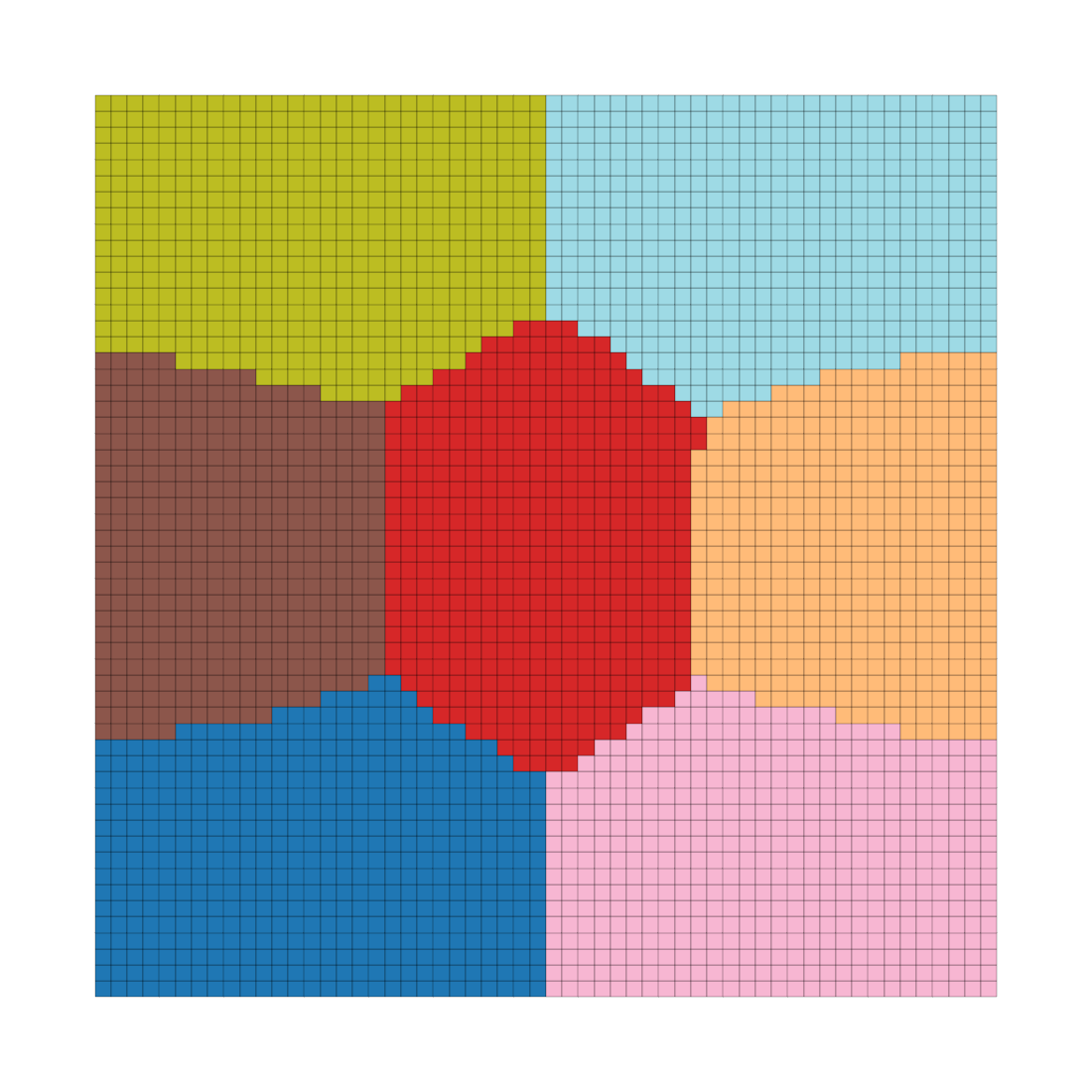}
            \caption{$56\times 56$ grid graph.}
        \end{subfigure} \hfill
        \begin{subfigure}[c]{0.53\linewidth}
            \centering
            \includegraphics[width=\linewidth]{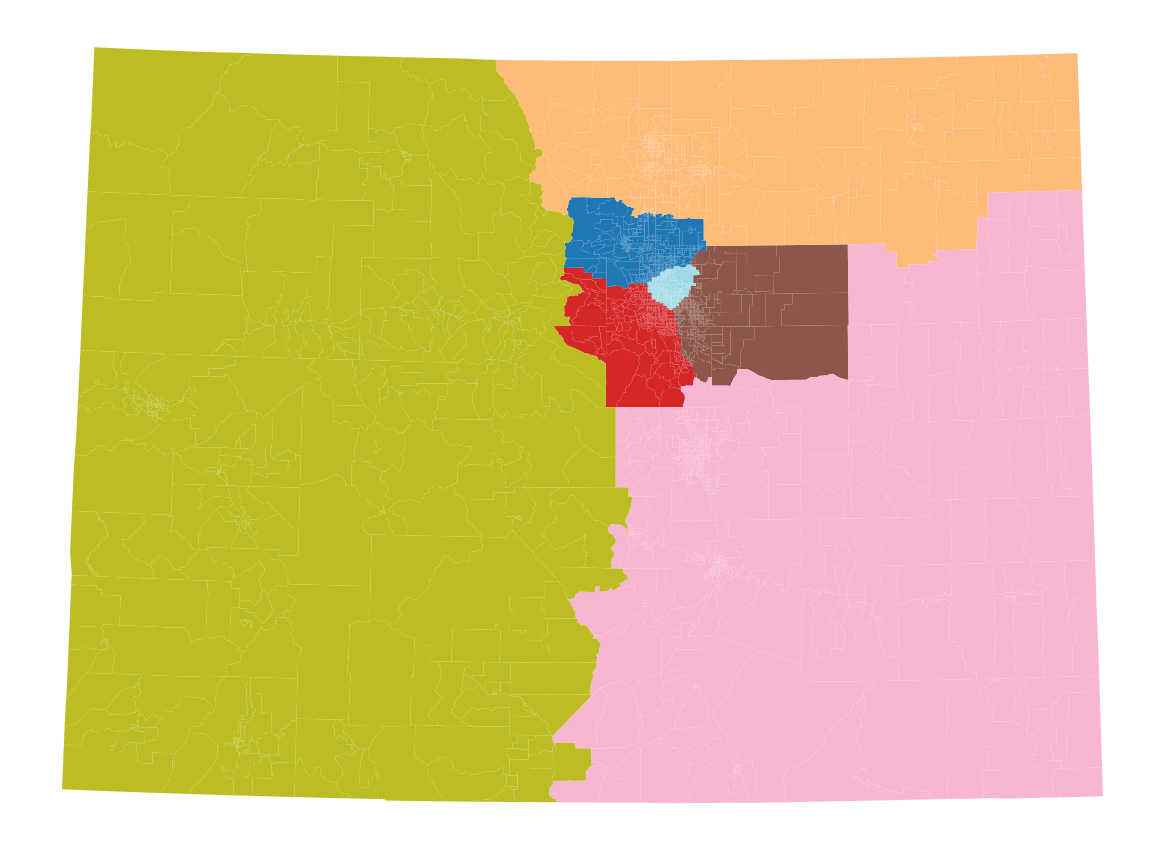}
            \caption{Colorado graph.}
        \end{subfigure}
        \caption{Example plans after 400 steps of SpecReCom.}
        \label{fig:specrecomplans}
    \end{figure}

    Figure~\ref{fig:specrecomplans} depicts two example connected $7$-partitions produced by SpecReCom after 400 steps of the chain. Visually, both plans are more ``compact'' than RevReCom's, and quantitatively, the parts/districts featured have fewer cut edges. 
    For the grid graph the starting state was the ``row'' partition seen in Figure~\ref{fig:initialgrid}, and for Colorado the starting state was given by the 2018 congressional districting plan.

    We note that SpecReCom does not take into account vertex weights. Therefore, while the plans may appear to be suitable for political use, their district populations are not necessarily well-balanced. For example, the partitions of the grid graph and Colorado graph in Figure~\ref{fig:specrecomplans} have population deviations of approximately 0.15 and 0.33 (respectively), both of which are worse than the majority of plans seen to be generated by RevReCom (see section~\ref{sec:results}).

    Preliminary work of a subset of the authors~\cite{summer_poster} attempted to incorporate vertex weight in SpecReCom by using a Laplacian with vertex weights, as described by Chung and Langlands \cite{chung_vertexlaplacian}. Unfortunately, this version of SpecReCom struggled to enforce connectedness, the Fiedler vector rarely helped find a connected bipartition of the recombined super-part. We hypothesize that including vertex weights in this way over-emphasized the importance of population balance at the expense of connectedness.

\subsection{Balanced Spectral Recombination}\label{sec:balspecrecom}
    Balanced Spectral Recombination (BalSpecReCom) attempts to resolve the weight balance issues of SpecReCom. In SpecReCom, the Fiedler cut is always made using a \emph{threshold} of zero in the sense that the bipartition of the graph $H$ induced by the recombined parts is
    \[
        \{\fiedler{H}^+, \fiedler{H}^-\} = \big\{\{v \in V_G : \fiedler{H}[v] \geq 0\},\ \{v \in V_G : \fiedler{H}[v] < 0 \} \big\}.
    \]
    The value zero on the right-hand side is the threshold that determines which side of the cut a vertex belongs to, and in general one can change the threshold to explore a set of spectrally-motivated cuts.
    In BalSpecReCom, we include a brute-force search for the threshold that optimizes for population balance, subject to the constraint that the parts are connected.
    
    Naively, one might worry that it is time-consuming to investigate all possible thresholds (since they are real-valued).
    It suffices, however, to consider just the entries of the Fiedler vector themselves as the threshold candidates, since all possible bipartitions which arise from some choice of threshold arise from these special thresholds.
    Algorithm~\ref{alg:balspecrecom} gives the implementation of the proposal.
    
    \begin{algorithm}[htbp]
        \caption{Balanced Spectral Recombination Proposal.}
        \begin{algorithmic}[1]
            \Procedure{BalSpecReCom}{$\D; G$} \Comment{input: current state; graph}
                \Select\ $uv \in \cutedges{G}{\D}$ UAR \Comment{sample a cut edge}
                \State $H \gets \inducedsubgraph{G}{\districtof{u} \cup \districtof{v}}$ \Comment{recombine}
                \For{$e \in E(H)$}
                    \State $w(e) \gets $ \Call{rand}{1,2} \Comment{randomize edge weights}
                \EndFor
                \State \textit{popDiff} $\gets \infty$
                \State \textit{splits} $\gets \{\{\districtof{u}, \districtof{v}\}\}$
                \For{$t \in \fiedler{H}$} \Comment{brute force search for optimal threshold(s)}
                    \State $\fiedler{H}^{\geq t} \gets \{v \in V_H : \fiedler{H}[v] \geq t\}$
                    \State $\fiedler{H}^{< t} \gets \{v \in V_H : \fiedler{H}[v] < t\}$
                    \State $h_1, h_2 \gets \inducedsubgraph{H}{\fiedler{H}^{\geq t}}, \inducedsubgraph{H}{\fiedler{H}^{< t}}$
                    \If{$h_1$ and $h_2$ are both connected graphs}\Comment{only allow connected partitions}
                        \If{$|w(h_1) - w(h_2)| \le$ \textit{popDiff}}
                            \State \textit{splits} $\gets \{\{h_1, h_2\}\}$
                            \State \textit{popDiff} $\gets |w(h_1) - w(h_2)|$
                        \EndIf
                    \EndIf
                \EndFor
                \State $H_1, H_2 \gets \underset{S \in \textit{splits}}{\mathrm{argmin}}\ |\cutedges{H}{S}|$ \Comment{break ties with cut edge count}
                \State $\D' \gets (\D \setminus \{\districtof{u}, \districtof{v}\}) \cup \{V(H_1), V(H_2)\}$ \Comment{split}
                \State \textbf{return} $\D'$ \Comment{output: next state}
            \EndProcedure
        \end{algorithmic}
        \label{alg:balspecrecom}
    \end{algorithm}

    The threshold search is performed for each part recombination, and it simply checks the difference in part weights induced by using each Fiedler vector entry as a threshold (lines 7-21 of Algorithm~\ref{alg:balspecrecom}). In the event that multiple entries result in optimal population balance, the entry among them that minimizes the number of cut edges is chosen (line 22).

\begin{figure}[htbp]
        \centering
        \begin{subfigure}[c]{0.42\linewidth}
            \centering
            \includegraphics[width=\linewidth]{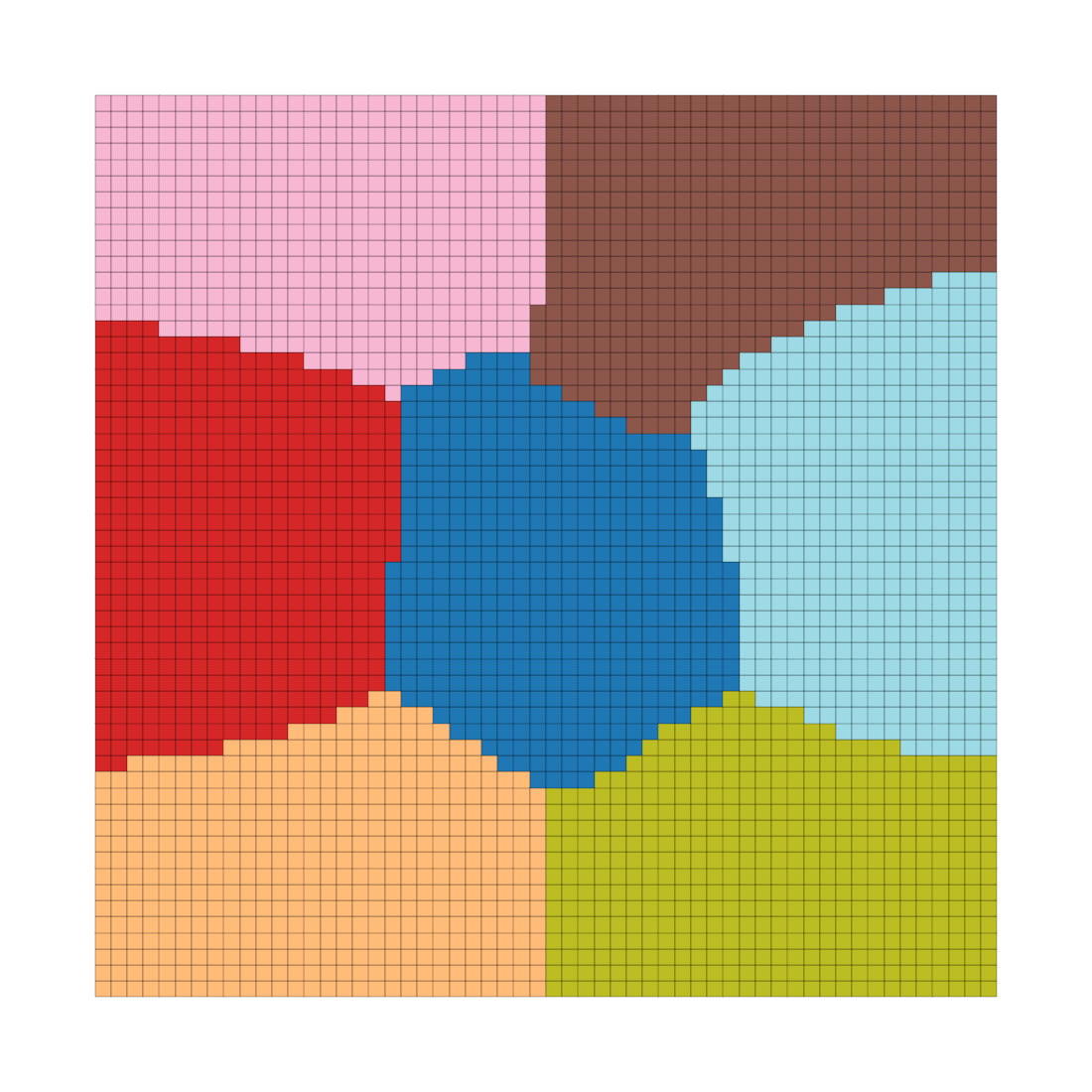}
            \caption{$56\times 56$ grid graph.}
        \end{subfigure} \hfill
        \begin{subfigure}[c]{0.53\linewidth}
            \centering
            \includegraphics[width=\linewidth]{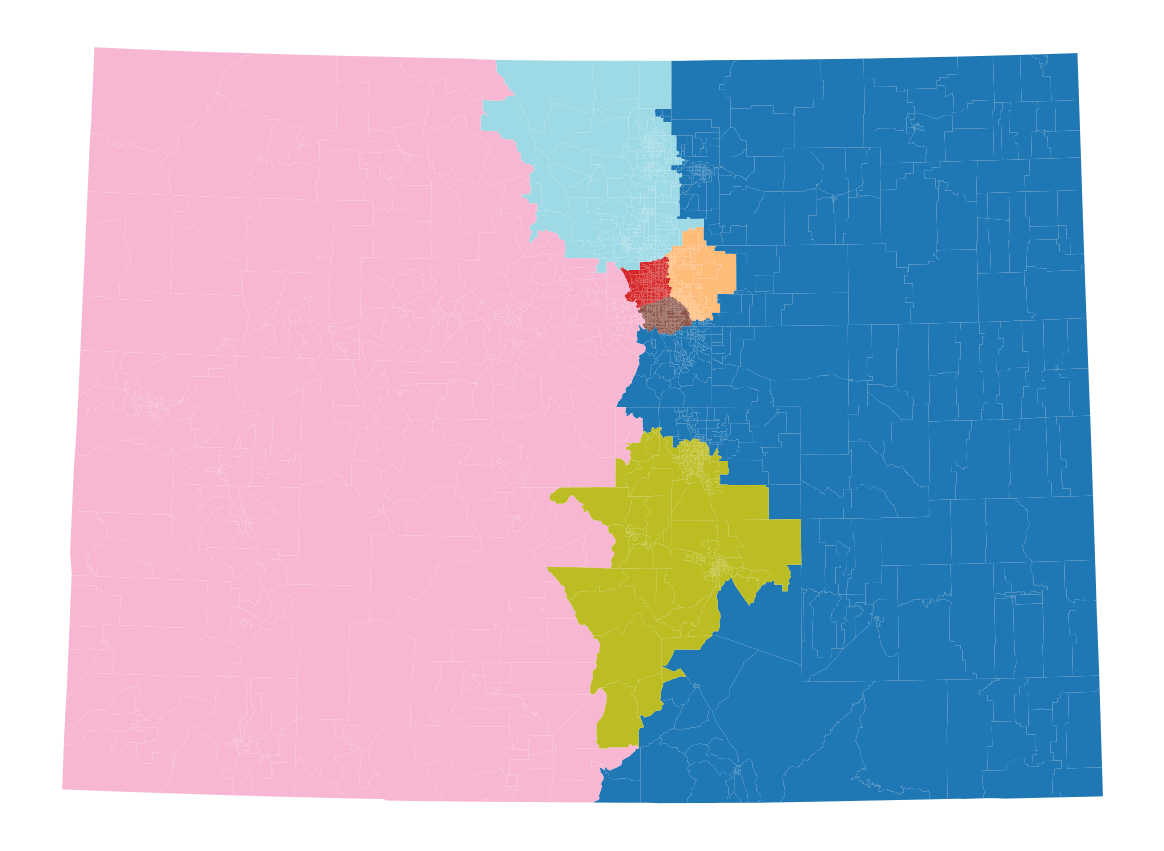}
            \caption{Colorado graph.}
        \end{subfigure}
        \caption{Example plans after 400 steps of BalSpecReCom.}
        \label{fig:balspecrecomplans}
    \end{figure}

    In Figure~\ref{fig:balspecrecomplans}, example plans generated by BalSpecReCom are shown. While the parts are less compact than those generated by SpecReCom, the population deviations are greatly improved. The grid graph plan in figure~\ref{fig:balspecrecomplans} has a perfect population deviation of 0.0, and the Colorado graph plan has a population deviation of approximately 0.002. This tradeoff is quantified in section~\ref{sec:results}.

\section{Implementation} \label{sec:implementation}

    Our implementations of SpecReCom and BalSpecReCom are both written in Python, and utilize the GerryChain library for political redistricting \cite{gerrychain}. We found that experimentation time could be reduced using a ``Redistricting'' Python package to contain experiment instances. Each instance holds all of the relevant data for a particular redistricting algorithm, ready to run with a set of tunable parameters, that automatically formats and outputs the necessary data at the end of the instance's run.
    
    Moreover, to provide a baseline with which to compare cut edge counts, we implemented a ``spectral $k$-means clustering'' (SpecKMeans) algorithm. This algorithm simply combines the idea of embedding vertices in a vector space, as done by Lee, Oveis Gharan and Trevisan~\cite{lee_cheegercut}, with $k$-means clustering to partition these embedded vertices into districts. This algorithm is deterministic and completely blind to the connectedness constraint, but serves as a good baseline by achieving low cut edge counts with few lines of code.
    
    All source code for the Redistricting package, as well as the data collected by it, is available at \url{https://github.com/MaxFlorescence/spectral_redistricting}.

\section{Evaluation} \label{sec:evaluation}

\subsection{Methodology}
    To evaluate whether SpecReCom and BalSpecReCom sample connected $k$-partitions with low cut edge counts, we experimentally compared the two to RevReCom's performance. We wrote several SLURM scripts to run many Redistricting instances on the CSU Computer Science department's Falcon cluster (\url{https://sna.cs.colostate.edu/hpc/}). 
    Our independent variables for each experiment were the algorithm that was run and the graph that it was run on. Our dependent variables were cut edge count and population deviation of the resulting plan. 
    The details are summarized in Table~\ref{tab:falcon}. This included running 10,000 SpecReCom and BalSpecReCom chains for 400 steps each, across 20 cores.
        
    \begin{table}[htbp]
        \centering
        \caption{Summary of experimental configurations.}
        \begin{tabular}{lllll}
        \toprule
            \textbf{Algorithm} & \textbf{Graph} & \textbf{Steps} ($\boldsymbol{N}$) & \textbf{Districts} ($\boldsymbol{k}$) & \textbf{Ensemble Size} ($\boldsymbol{m}$) \\ 
        \midrule
            \multirow{2}{*}{RevReCom} & 56$\times$56 grid graph & \multirow{2}{*}{10,000} & \multirow{2}{*}{7} & \multirow{2}{*}{10,000} \\
        \cmidrule{2-2}
            & 2018 Colorado districts & & & \\
        \cmidrule{1-5}
            \multirow{2}{*}{SpecReCom} & 56$\times$56 grid graph & \multirow{2}{*}{400} & \multirow{2}{*}{7} & \multirow{2}{*}{10,000} \\
        \cmidrule{2-2}
            & 2018 Colorado districts & & & \\
        \cmidrule{1-5}
            \multirow{2}{*}{BalSpecReCom} & 56$\times$56 grid graph & \multirow{2}{*}{400} & \multirow{2}{*}{7} & \multirow{2}{*}{10,000} \\
        \cmidrule{2-2}
            & 2018 Colorado districts & & & \\
        \bottomrule
        \end{tabular}
        \label{tab:falcon}
    \end{table}
    
    While GerryChain provides an implementation of RevReCom, we opted to use an optimized Rust implementation written by Rule et al.\ \cite{rust_revrecom} to generate many different plans. Then to facilitate data collection, we used a technique known as sub-sampling, as described in \textit{Political Geometry} \cite{duchin_politicalgeometry}. Instead of running $m$ chains independently for $N$ steps each, we run a single chain for $m \cdot N$ steps, sampling one state every $N$ steps to produce $m$ plans. In our case, $m = N =$ 10,000.

    To provide a baseline with which to compare our results from the Colorado graph, we chose to use a square grid graph with 56 vertices per side. This value was chosen as it was the integer whose square ($56^2 = 3136$) was the closest to the number of vertices in the Colorado graph ($3135$). Coincidentally, as 56 is a multiple of 7, a $56 \times 56$ grid graph admits $7$-partitions with equal-sized parts. 

\subsection{Results} \label{sec:results}
\begin{figure}[htbp]
        \centering
        \begin{subfigure}[c]{0.34\linewidth}
            \centering
            \includegraphics[width=\linewidth]{recom_1.png}
            \caption{$56\times 56$ grid graph.}
        \end{subfigure} \hfill
        \begin{subfigure}[c]{0.53\linewidth}
            \centering
            \includegraphics[width=\linewidth]{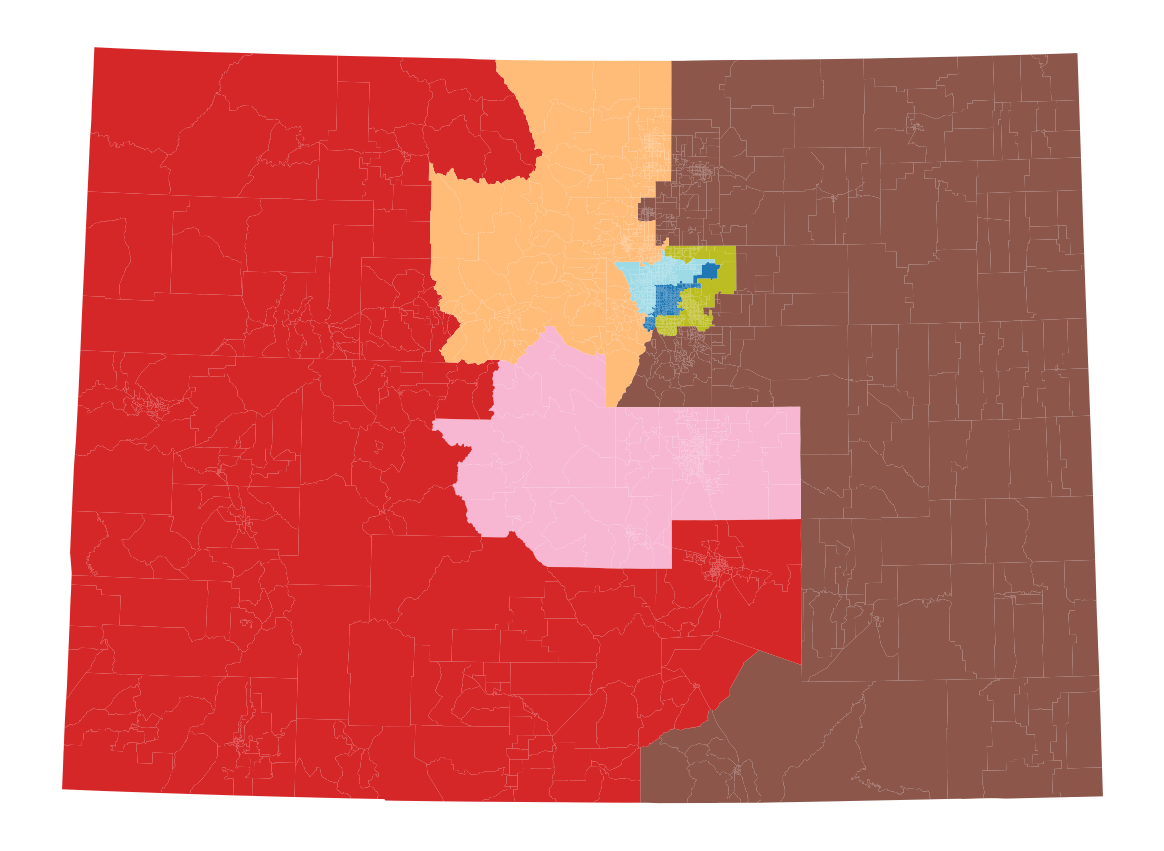}
            \caption{Colorado graph.}
        \end{subfigure}
        \caption{Initial plans for redistricting chains.}
        \label{fig:initial_plans}
    \end{figure}

\begin{figure}[htbp]
        \centering
        \begin{subfigure}[c]{0.47\linewidth}
            \includegraphics[width=\linewidth]{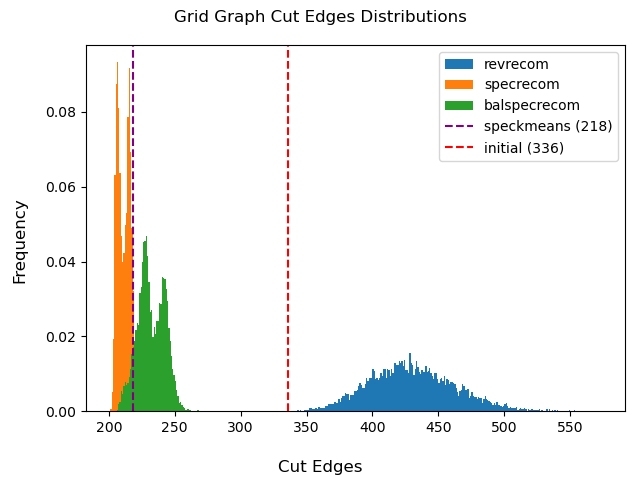}
            \caption{$56\times 56$ grid graph.\label{fig:56x56gridresults}}
        \end{subfigure} \hfill
        \begin{subfigure}[c]{0.47\linewidth}
            \includegraphics[width=\linewidth]{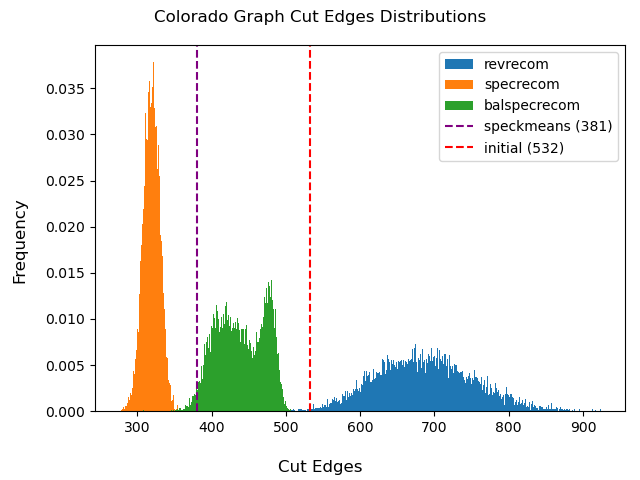}
            \caption{Colorado graph.}
        \end{subfigure}
        \caption{Cut edge distributions across each algorithm.}
        \label{fig:cutedgehist}
    \end{figure}

\begin{figure}[htbp]
        \centering
        \begin{subfigure}[c]{0.47\linewidth}
            \includegraphics[width=\linewidth]{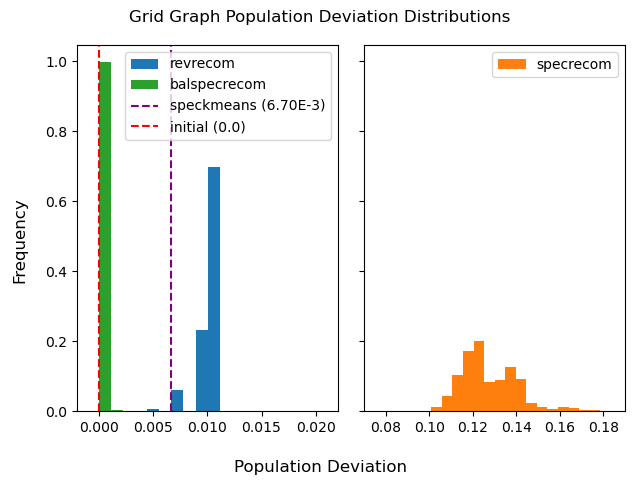}
            \caption{$56\times 56$ grid graph.}
        \end{subfigure} \hfill
        \begin{subfigure}[c]{0.47\linewidth}
            \includegraphics[width=\linewidth]{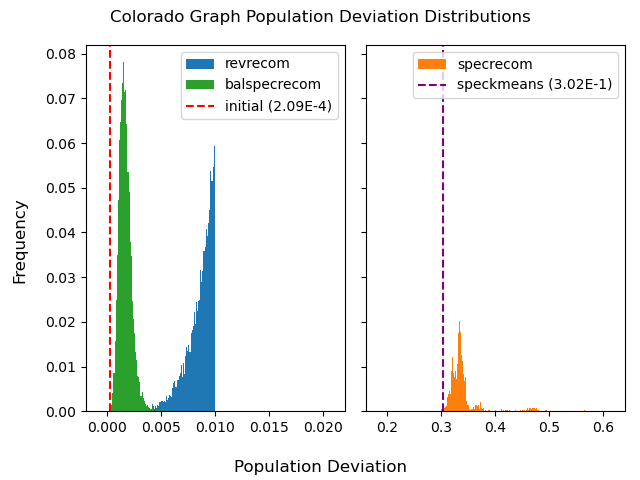}
            \caption{Colorado graph.}
        \end{subfigure}
        \caption{Population deviation distributions across each algorithm.}
        \label{fig:popdisphist}
    \end{figure}

    Figures~\ref{fig:cutedgehist} and \ref{fig:popdisphist} show the distribution of cut edge counts and population deviations respectively, across the 6 different combinations of algorithm and graph. Each histogram is normalized independently. In orange is the data for RevReCom, in blue is SpecReCom, and in green in BalSpecReCom. A vertical, dashed red line indicates the initial values of the graph for each metric. A vertical, dashed purple line indicates the baseline SpecKMeans value. For the $56\times 56$ grid graph, the initial $7$-partition was as in Figure~\ref{fig:initialgrid}, and for the Colorado graph the initial $7$-partition was the 2018 congressional district map. (See Figure~\ref{fig:initial_plans}.)

\subsection{Discussion} \label{sec:discussion}
    In terms of cut edges, we see a clear difference between our SpecReCom and BalSpecReCom algorithms and RevReCom. Figure~\ref{fig:cutedgehist} shows that for both the $56\times 56$ grid graph and the Colorado graph, SpecReCom achieves the lowest numbers of cut edges in its final generated $7$-partitions, followed closely by BalSpecReCom plans, and then RevReCom plans with around twice as many cut edges on average. RevReCom's distribution also lies mostly above the initial cut edges value for each graph, while SpecReCom and BalSpecReCom lie below it. Relative to the SpecKMeans baseline, only SpecReCom manages to mostly generate plans with fewer cut edges.

    In terms of population deviation, the large difference between SpecReCom and BalSpecReCom or RevReCom necessitated that the plot be split. Figure~\ref{fig:popdisphist} shows that BalSpecReCom and RevReCom both manage to get population deviation values below 0.01 (a 1\% tolerance), while SpecReCom achieves values for the Colorado graph that are an order of magnitude higher. For the $56\times 56$ grid graph, BalSpecReCom most frequently manages to maintain the perfect population balance of the initial plan, and for the Colorado graph it consistently achieves deviations less than $0.005$. In both cases, SpecReCom achieved values mostly above the SpecKMeans baseline, and BalSpecReCom achieved values below it.

    In the case of RevReCom, the sharp cutoff at 0.01 population deviation can be attributed to the algorithm's implementation. Using $\varepsilon$-balance edges, RevReCom never accepts $7$-partitions that exceed a population deviation of $\varepsilon$. For the implementation of RevReCom that we used, $\varepsilon = 0.01$.
    
\section{Conclusions} \label{sec:conclusions}
    Our results provide empirical evidence in favor of the hypothesis that Fiedler vectors can be used to create connected $k$-partitions with much shorter boundaries than existing algorithms from the literature.
    Figure~\ref{fig:cutedgehist} demonstrates that SpecReCom and BalSpecReCom both tend to generate these partitions with fewer cut edges than RevReCom, and in much fewer steps (400 as opposed to 10,000). Both the $56\times 56$ grid graph and the Colorado graph exhibit this behavior, indicating that spectral recombination methods can work across different types of graphs. Moreover, SpecReCom produces $k$-partitions with fewer cut edges and a higher affinity for connected parts than a simple clustering approach like SpecKMeans. That is, SpecReCom is more likely to produce \emph{connected} $k$-partitions, whereas SpecKMeans would need to be modified to guarantee connected parts.

    Figure~\ref{fig:popdisphist} demonstrates that the SpecReCom algorithm can be used as a base to create new algorithms that effectively take different redistricting metrics into consideration, specifically population deviation. RevReCom achieves population deviation of at most 0.01, since this is the value of $\varepsilon$ used in our experimentation. However, due to SpecReCom making no attempt to maintain population balance within the same tolerance $\varepsilon$, it achieves significantly higher population deviation values for both graphs. In contrast, BalSpecReCom mostly achieves population deviation values lower than those achieved by RevReCom, owing to its brute-force search for the optimal spectral bipartitioning threshold (with respect to population deviation) during each step. The values given by the SpecKMeans also reveal that a smarter algorithm is necessary; while it does achieve a population deviation below 0.01 for the $56\times 56$ grid graph, it is on-par with SpecReCom for the real-world Colorado graph.

    When taken together, these results imply that there exists a tradeoff between generating districting connected $k$-partitions with few cut edges and with low population deviation. Figure~\ref{fig:popdisphist} clearly shows that BalSpecReCom generates these partitions with population deviations lower than SpecReCom, but \ref{fig:cutedgehist} indicates that this comes at the cost of increasing the number of cut edges relative to SpecReCom.
    
\section{Future Work} \label{sec:future}

DeFord et al.\ \cite{deford_recomfamily} provide several open questions, one of which is, ``Propose other balanced bipartitioning methods to replace spanning trees, supported by fast algorithms''. Additionally, an invitation written by DeFord and Duchin \cite{duchin_politicalgeometry} reads, ``New methods should strive for easy implementation, low rejection rate, adaptability to varied districting criteria, and of course theoretical properties like provable ergodicity or reversibility''.

In this section, we list several areas in which our work can be expanded. The first two relate to improving our algorithms' robustness, the next two to fully addressing the open questions mentioned above, and the last to what work needs to be done before applying our algorithms to redistricting in the real world.

\subsection{Alternate Bipartitioning Vectors} \label{sec:alternatefiedler}
    As Urschel and Zikatanov note~\cite{urschel_bisection}, there may be more than one Fiedler vector for a particular graph, depending on the geometric multiplicity of the second eigenvalue of the Laplacian. Moreover, Colin de Verdi\`ere's invariant \cite{verdire_multiplicity} reveals that the corank (and thus the geometric multiplicity of the algebraic connectivity) of a planar graph's Laplacian can be at most three. Therefore, even though our spectral recombination algorithms seem to be well-suited for producing ensembles, we do not thoroughly explore the full eigenspaces. Future work might consider an optimization over the entire eigenspace of Fiedler vectors.

\subsection{Improved Balancing Sweep}
    Since our BalSpecReCom algorithm considers applying a variable threshold to the Fiedler vector's entries, it strictly enforces the order of the graph's vertices according to those entries. This may not be ideal, as the optimal split in terms of population deviation might be inaccessible since it could require a slightly different ordering. To this end, future work might consider some kind of ``noisy sweep'', in which for each threshold the vertices near the threshold can be permuted. Considering all permutations would lead to an exponential increase in running time, and so perhaps a single, randomly chosen permutation would be considered instead. 

    Alternatively, we note that Recom and RevReCom take a weight-balancing parameter, $\varepsilon$, while BalSpecReCom only considers the threshold that maximizes weight balance at each step. Instead, BalSpecReCom could consider ``$\varepsilon$-balance thresholds''---any threshold that achieves weight balance within some $\varepsilon$ tolerance. This modification may allow plans with fewer cut edges \emph{and} acceptable population deviation to be generated, addressing the tradeoff discussed in Section~\ref{sec:discussion}.

\subsection{Algorithm Optimization}
    Recall that the open question of DeFord et al.\ asks researchers to ``propose other balanced bipartitioning methods to replace spanning trees, \emph{supported by fast algorithms}'' \cite{deford_recomfamily}. Regarding this, we made no attempt to optimize SpecReCom or BalSpecReCom beyond basic implementations in Python.
    
    A Rust implementation of RevReCom by Rule et al.\ \cite{rust_revrecom} was optimized to perform steps of the chain in parallel, decreasing the amount of time the chain spends rejecting possible moves. In our experimentation, we found that with 8 cores the Rust implementation of RevReCom completes approximately 55.3 thousand steps per second on the Colorado graph. While our implementations of SpecReCom and BalSpecReCom are not comparable to this implementation of RevReCom, we have observed practically that they take much more time to generate the same number of plans.
    
    Given that the majority of SpecReCom's computational load is spent performing matrix operations, we hypothesize that optimizing it would be relatively easy. However, BalSpecReCom includes all of SpecReCom's code as well as a brute force loop through the Fiedler vector, adding an $O(n)$ factor to the time complexity. An optimized version might consider parallelizing this loop or replacing it with a smarter method of searching for an acceptable threshold, such as stopping early once a single $\varepsilon$-balance threshold is found.

\subsection{Statistical and Theoretical Analysis}
     Recall that DeFord and Duchin's invitation asks for ``theoretical properties like provable ergodicity and reversibility'' \cite{duchin_politicalgeometry}. While our work demonstrates that \emph{spectral} recombination algorithms are viable, we do not know if either SpecReCom or BalSpecReCom is aperiodic or irreducible, and reversibility is likely false. Future work could give insight in this area, providing a rigorous understanding of the chains by determining their stationary distributions and (in particular) mixing times.
    
    In the absence of theoretical guarantees, Chikina et al.\ describe several statistical tests for whether a chain has mixed~\cite{chikina_mixedtest}. The ``$\sqrt{\varepsilon}$'' test would allow it to be determined with significance $p = \sqrt{2\varepsilon}$ whether or not a given run of a Markov chain is mixed at its current state. Furthermore, an extension of this test allows for the effect size $\varepsilon$ to be separated from the significance $p$ \cite{chikina_unusualtest}.

\subsection{Considering Additional Plan Constraints}
    Clelland et al.\ \cite{clelland_coloradoanalysis} list districting plan requirements for Colorado, two of which are that plans must ``preserve communities of interest and political subdivisions'' and ``minimize the number of divisions when a city, county, or town is divided''.
    
    While our algorithms do not consider these constraints during the chains' runs, we note that GerryChain does provide methods for determining how many units are split by a given plan. Moreover, these methods are parameterized by which type of unit (counties, precincts, etc.) is being considered. Future work could analyze how SpecReCom and BalSpecReCom perform under these metrics, and propose modifications to comply with them, if necessary.
    Another constraint listed by Clelland et al.\ is that plans must ``maximize the number of politically competitive districts''. Again, GerryChain provides metrics for determining how competitive a given districting plan is, and so the same can be said about future work with regards to this constraint.
    
    We have demonstrated that our SpecReCom can easily be modified to account for population deviation, yielding BalSpecReCom. Future work might be able to modify SpecReCom in similar ways to account for the metrics and constraints discussed above.

\bibliographystyle{habbrv}
\bibliography{bib}

\begin{thebibliography}{10}
\expandafter\ifx\csname url\endcsname\relax
  \def\url#1{\texttt{#1}}\fi
\expandafter\ifx\csname doi\endcsname\relax
  \def\doi#1{\burlalt{\textsc{doi}:\detokenize{#1}}{https://dx.doi.org/#1}}\fi
\expandafter\ifx\csname
  urlprefix\endcsname\relax\def\urlprefix{\textsc{url:}}\fi
\expandafter\ifx\csname href\endcsname\relax
  \def\href#1#2{#2}\fi
\expandafter\ifx\csname burlalt\endcsname\relax
  \def\burlalt#1#2{\href{#2}{#1}}\fi

\bibitem{Alo86}
N.~Alon.
\newblock Eigenvalues and expanders.
\newblock {\em Combinatorica}, 6(2):83--96, 1986.
\newblock \doi{10.1007/BF02579166}.

\bibitem{AM85}
N.~Alon and V.~D. Milman.
\newblock {$\lambda_1$}, {{Isoperimetric Inequalities for Graphs, and
  Superconcentrators}}.
\newblock {\em Journal of Combinatorial Theory, Series B}, 38(1):73--88, 1985.
\newblock \doi{10.1016/0095-8956(85)90092-9}.

\bibitem{gerrychain}
D.~and Democracy~Lab.
\newblock Gerrychain.
\newblock Python Library, 2018.
\newblock \urlprefix\url{https://github.com/mggg/GerryChain}.

\bibitem{cannon_revrecom}
S.~Cannon, M.~Duchin, D.~Randall, and P.~Rule.
\newblock Spanning tree methods for sampling graph partitions.
\newblock 2022, \burlalt{arXiv:2210.01401}{https://arxiv.org/abs/2210.01401}.

\bibitem{chen_detectgerrymander}
J.~Chen and J.~Rodden.
\newblock Cutting through the thicket: Redistricting simulations and the
  detection of partisan gerrymanders.
\newblock {\em Election Law Journal: Rules, Politics, and Policy},
  14(4):331--345, 2015.
\newblock \doi{10.1089/elj.2015.0317}.

\bibitem{chikina_unusualtest}
M.~Chikina, A.~Frieze, J.~C. Mattingly, and W.~Pegden.
\newblock {Separating Effect From Significance in Markov Chain Tests}.
\newblock {\em Statistics and Public Policy}, 7(1):101--114, 2020.
\newblock \doi{10.1080/2330443X.2020.1806763}.

\bibitem{chikina_mixedtest}
M.~Chikina, A.~Frieze, and W.~Pegden.
\newblock {Assessing significance in a Markov chain without mixing}.
\newblock {\em Proceedings of the National Academy of Sciences},
  114(11):2860--2864, Mar. 2017.
\newblock \doi{10.1073/pnas.1617540114}.

\bibitem{chung_sgt}
F.~R.~K. Chung.
\newblock {\em Spectral graph theory}.
\newblock American Mathematical Society, 1997.

\bibitem{chung_vertexlaplacian}
F.~R.~K. Chung and R.~P. Langlands.
\newblock {A Combinatorial Laplacian with Vertex Weights}.
\newblock {\em Journal of Combinatorial Theory, Series A}, 75(2):316--327,
  1996.
\newblock \doi{10.1006/jcta.1996.0080}.

\bibitem{clelland_coloradoanalysis}
J.~Clelland, H.~Colgate, D.~DeFord, B.~Malmskog, and F.~Sancier-Barbosa.
\newblock {Colorado in context: Congressional redistricting and competing
  fairness criteria in Colorado}.
\newblock {\em Journal of Computational Social Science}, 5(1):189--226, 2022.
\newblock \doi{10.1007/s42001-021-00119-7}.

\bibitem{clelland_recom}
J.~N. Clelland, N.~Bossenbroek, T.~Heckmaster, A.~Nelson, P.~Rock, and
  J.~VanAusdall.
\newblock Compactness statistics for spanning tree recombination.
\newblock 2021, \burlalt{arXiv:2103.02699}{https://arxiv.org/abs/2103.02699}.

\bibitem{cohenaddad_polygondistricts}
V.~{Cohen-Addad}, P.~N. Klein, and N.~E. Young.
\newblock Balanced centroidal power diagrams for redistricting.
\newblock In {\em Proceedings of the 26th {{ACM SIGSPATIAL International
  Conference}} on {{Advances}} in {{Geographic Information Systems}}},
  {{SIGSPATIAL}} '18, pages 389--396, New York, NY, USA, 2018. Association for
  Computing Machinery.
\newblock \doi{10.1145/3274895.3274979}.

\bibitem{summer_poster}
E.~Davies, R.~Job, H.~Kim, and H.~Seo.
\newblock A spectral approach to legislative districting in colorado.
\newblock Research poster, Summer 2023.

\bibitem{verdire_multiplicity}
Y.~C. de~Verdi{\`e}re.
\newblock On a novel graph invariant and a planarity criterion.
\newblock {\em J. Comb. Theory Ser.\ B}, 50(1):11--21, Aug. 1990.
\newblock \doi{10.1016/0095-8956(90)90093-F}.

\bibitem{deford_personal}
D.~DeFord.
\newblock Gerryprojects, 2023.

\bibitem{deford_recomfamily}
D.~DeFord, M.~Duchin, and J.~Solomon.
\newblock Recombination: A family of markov chains for redistricting.
\newblock {\em Harvard Data Science Review}, 2021.
\newblock \doi{10.1162/99608f92.eb30390f}.

\bibitem{duchin_politicalgeometry}
M.~Duchin and O.~Walch.
\newblock {\em Political geometry: Rethinking redistricting in the US with
  math, law, and everything in between}.
\newblock Birkh\"auser, 2022.

\bibitem{fiedler_connectivity}
M.~Fiedler.
\newblock Algebraic connectivity of graphs ({E}nglish).
\newblock {\em Czechoslovak Mathematical Journal}, 23, 1973.
\newblock \doi{10.21136/CMJ.1973.101168}.

\bibitem{fifield_firstmcmc}
B.~Fifield, M.~Higgins, K.~Imai, and A.~Tarr.
\newblock Automated redistricting simulation using markov chain monte carlo.
\newblock {\em Journal of Computational and Graphical Statistics},
  29(4):715--728, 2020.
\newblock \doi{10.1080/10618600.2020.1739532}.

\bibitem{For64}
E.~Forrest.
\newblock Apportionment by computer.
\newblock {\em The American Behavioral Scientist (pre-1986)}, 8(4):23, 1964.
\newblock \doi{10.1177/000276426400800407}.

\bibitem{GJ09}
M.~R. Garey and D.~S. Johnson.
\newblock {\em Computers and intractability: a guide to the theory of
  NP-completeness}.
\newblock Freeman, New York, 2009.

\bibitem{mggg_techreport}
M.~Geometry and G.~Group.
\newblock Comparison of districting plans for the virginia house of delegates.
\newblock Technical report, Cornell University, Brooks School of Public Policy,
  2018.
\newblock \urlprefix\url{https://mggg.org/VA-report.pdf}.

\bibitem{herschlag_applyensembles}
G.~Herschlag, H.~S. Kang, J.~Luo, C.~V. Graves, S.~Bangia, R.~Ravier, and J.~C.
  Mattingly.
\newblock {Quantifying Gerrymandering in North Carolina}.
\newblock {\em Statistics and Public Policy}, 7(1):30--38, 2020.
\newblock \doi{10.1080/2330443X.2020.1796400}.

\bibitem{Kar93}
D.~R. Karger.
\newblock Global min-cuts in {RNC}, and other ramifications of a simple min-cut
  algorithm.
\newblock In {\em Proceedings of the fourth annual ACM-SIAM symposium on
  Discrete algorithms}, SODA '93, page 21–30, USA, 1993. Society for
  Industrial and Applied Mathematics.

\bibitem{KS96a}
D.~R. Karger and C.~Stein.
\newblock A new approach to the minimum cut problem.
\newblock {\em Journal of the ACM}, 43(4):601–640, 1996.
\newblock \doi{10.1145/234533.234534}.

\bibitem{kawahara_disparity}
J.~Kawahara, T.~Horiyama, K.~Hotta, and {\relax Si}.~Minato.
\newblock Generating all patterns of graph partitions within a disparity bound.
\newblock In S.-H. Poon, M.~S. Rahman, and H.-C. Yen, editors, {\em WALCOM:
  Algorithms and Computation}, pages 119--131, Cham, 2017. Springer
  International Publishing.
\newblock \doi{10.1007/978-3-319-53925-6_10}.

\bibitem{KLL+13}
T.~C. Kwok, L.~C. Lau, Y.~T. Lee, S.~Oveis~Gharan, and L.~Trevisan.
\newblock Improved {C}heeger's inequality: analysis of spectral partitioning
  algorithms through higher order spectral gap.
\newblock In {\em Proceedings of the 45th annual ACM symposium on Symposium on
  theory of computing - STOC '13}, page~11, Palo Alto, California, USA, 2013.
  ACM Press.
\newblock \doi{10.1145/2488608.2488611}.

\bibitem{lee_cheegercut}
J.~R. Lee, S.~O. Gharan, and L.~Trevisan.
\newblock Multiway spectral partitioning and higher-order cheeger inequalities.
\newblock {\em J. ACM}, 61(6), 2014.
\newblock \doi{10.1145/2665063}.

\bibitem{levin_voronoi}
H.~A. Levin and S.~A. Friedler.
\newblock Automated congressional redistricting.
\newblock {\em ACM J. Exp. Algorithmics}, 24, Apr. 2019.
\newblock \doi{10.1145/3316513}.
\newblock \url{https://doi.org/10.1145/3316513}.

\bibitem{OT12}
S.~Oveis~Gharan and L.~Trevisan.
\newblock Approximating the expansion profile and almost optimal local graph
  clustering.
\newblock In {\em 2012 IEEE 53rd Annual Symposium on Foundations of Computer
  Science}, pages 187--196, New Brunswick, NJ, USA, 2012. IEEE.
\newblock \doi{10.1109/FOCS.2012.85}.

\bibitem{OT14}
S.~Oveis~Gharan and L.~Trevisan.
\newblock Partitioning into expanders.
\newblock In {\em Proceedings of the Twenty-Fifth Annual ACM-SIAM Symposium on
  Discrete Algorithms}, pages 1256--1266. Society for Industrial and Applied
  Mathematics, Jan. 2014.
\newblock \doi{10.1137/1.9781611973402.93}.

\bibitem{rust_revrecom}
P.~J. Rule, M.~Sarahan, and M.~Fan.
\newblock Fastest recom chain in the west.
\newblock GitHub, 2022.
\newblock \urlprefix\url{https://github.com/pjrule/frcw.rs}.

\bibitem{SJ89}
A.~Sinclair and M.~Jerrum.
\newblock Approximate counting, uniform generation and rapidly mixing markov
  chains.
\newblock {\em Information and Computation}, 82(1):93--133, 1989.
\newblock \doi{10.1016/0890-5401(89)90067-9}.

\bibitem{urschel_bisection}
J.~C. Urschel and L.~T. Zikatanov.
\newblock Spectral bisection of graphs and connectedness.
\newblock {\em Linear Algebra and its Applications}, 449:1--16, 2014.
\newblock \doi{10.1016/j.laa.2014.02.007}.

\bibitem{vickery_gerrymandering}
W.~Vickrey.
\newblock On the prevention of gerrymandering.
\newblock {\em Political Science Quarterly}, 76(1):105--110, 1961.
\newblock \doi{10.2307/2145973}.

\end{thebibliography}

\end{document}